\documentclass[
  aps,
  prd,
  reprint,
  preprintnumbers,
  amsmath,
  amssymb,
  nofootinbib,
  floatfix,
  longbibliography
]{revtex4-2}

\usepackage[T1]{fontenc}

\usepackage{microtype}
\usepackage{graphicx}
\usepackage{booktabs}
\usepackage{multirow}
\usepackage{siunitx}
\usepackage{capt-of}
\usepackage{hyperref}

\DeclareMathOperator{\Res}{Res}
\newcommand{\dd}{\mathrm{d}}

\begin{document}

\preprint{OUTP-26-14P}

\title{Automated two-loop fermion-bubble contributions at high multiplicity}

\author{Dario Kermanschah}
\email{d.kermanschah@gmail.com}
\affiliation{Rudolf Peierls Centre for Theoretical Physics, University of Oxford, Clarendon Laboratory, Parks Road, Oxford OX1 3PU, United Kingdom}
\affiliation{Institute for Theoretical Physics, ETH Zurich, Wolfgang-Pauli-Strasse 27, 8093 Z\"urich, Switzerland}

\begin{abstract}
We compute the gauge-invariant, helicity-summed two-loop \(N_f\) contribution for \(q\bar q\to n\gamma^{(*)}\), \(n=2,\ldots,8\).
Local ultraviolet renormalisation, infrared factorisation, analytic loop-energy integration and a new automated threshold subtraction yield finite integrals over real spatial momenta.
Adaptive multi-channel Monte Carlo samples loop momenta, helicities and identical-particle permutations.
We quantify generation and evaluation costs across multiplicities.
The same representation is integrated directly with parton distributions through five photons, yielding inclusive and differential virtual corrections.
\end{abstract}

\maketitle

\section{Introduction}
\label{sec:introduction}

Two-loop virtual corrections are an essential ingredient of precision collider predictions and have been the last missing contribution in several recent next-to-next-to-leading-order (NNLO) calculations.
Their automated evaluation for broad classes of processes at increasing multiplicity remains a major goal.
Here we present an automated calculation of the gauge-invariant fermion-bubble \(N_f\) contribution to the two-loop hard finite remainder in quantum chromodynamics (QCD) for
\begin{equation}
q\bar q\longrightarrow n\gamma^{(*)},
\qquad n=2,\ldots,8,
\label{eq:processes-new2}
\end{equation}
for on-shell and equal-mass off-shell photons, reaching ten external legs.
The $N_f$ contribution provides a controlled setting in which to test the entire construction up to high multiplicities.
The second loop is a massless-fermion vacuum-polarisation insertion, such that the amplitude contains ultraviolet (UV) subdivergences and overlapping infrared (IR) singularities, while the physical thresholds remain of one-loop type.
\par

The traditional route to these amplitudes is to construct global analytic expressions in terms of master integrals and kinematic coefficients in dimensional regularisation.
Their substantial construction cost can be repaid by fast evaluation throughout phase space, as demonstrated for two-loop three-photon amplitudes~\mbox{\cite{Abreu:2020cwb,Chawdhry:2020for,Abreu:2023bdp}}.
However, the reduction systems, master integrals, and coefficients grow rapidly in complexity with the number of external particles and kinematic scales, which can make symbolic treatment impractical.
This has become particularly apparent for massive $2\to3$ processes, where numerical methods enter the construction, such as pointwise integral reduction, finite-field reconstruction, and numerical master-integral evaluation~\mbox{\cite{Hartanto:2019uvl,Badger:2021nhg,Abreu:2021asb,Agarwal:2024jyq,Badger:2024sqv,Badger:2024mir,Badger:2024dxo,DeLaurentis:2025dxw,Badger:2025ljy,DeLaurentis:2026brm,Becchetti:2026yxl}}.
These advances extend the method's reach, but the intended payoff of an efficient global expression is not always realised, and evaluation may still require precomputed phase-space grids and interpolation.
\par

Here we pursue a different organisation: we retain analytic control locally to obtain a finite integrand and use numerical methods only for its final integration.
UV divergences are locally subtracted using the \(R\)-operation~\mbox{\cite{Bogoliubov:1957gp,Hepp:1966eg,Zimmermann:1969jj,Gomes:1974cr,Caswell:1981ek,Herzog:2017jgk,Capatti:2022tit}} and IR divergences are cancelled by local counterterms~\mbox{\cite{Nagy:2003qn,Assadsolimani:2009cz,Anastasiou:2020sdt,Anastasiou:2022eym,Anastasiou:2024xvk,Anastasiou:2025cvy,Haindl:2025jte}} guided by factorisation~\mbox{\cite{Sterman:1978bi,Libby:1978qf,Akhoury:1978vq,Sen:1982bt,Collins:1989bt,Sterman:2002qn,Feige:2014wja,Ma:2019hjq}}.
After these subtractions, analytic loop-energy integration yields integrands algebraically identical to those of loop-tree duality and its causal formulations, which are related to time-ordered perturbation theory~\mbox{\cite{Kilian:2009wy,Catani:2008xa,Aguilera-Verdugo:2020set,JesusAguilera-Verdugo:2020fsn,Aguilera-Verdugo:2019kbz,Aguilera-Verdugo:2020kzc,Runkel:2019yrs,Capatti:2019ypt,Capatti:2020ytd,Ramirez-Uribe:2020hes,Ramirez-Uribe:2022sja,Sborlini:2021owe,TorresBobadilla:2021ivx,Benincasa:2021qcb,Driencourt-Mangin:2019aix,Rios-Sanchez:2024xtv,Capatti:2022mly,Bodwin:1984hc,Sterman:2023xdj,Kermanschah:2026dzb}}.
The remaining threshold singularities are subtracted before adaptive multi-channel Monte Carlo integration \mbox{\cite{Kermanschah:2021wbk,Kermanschah:2024utt,Vicini:2024ecf,Kermanschah:2025wlo,Kilian:2009wy}}.
The central problem is therefore shifted from solving the loop correction globally to making its integrand locally integrable.
\par

Direct numerical loop integration has already demonstrated its viability for high-multiplicity amplitudes at one loop.
Refs.~\mbox{\cite{Becker:2011vg,Becker:2012aqa}} used local IR and UV subtraction to obtain leading-colour $e^+e^-$ jet rates through seven jets at next-to-leading order (NLO).
They rely on contour deformation~\mbox{\cite{Gong:2008ww,Becker:2012bi,Nagy:2006xy,Becker:2010ng,Soper:1999xk,Capatti:2019edf,Buchta:2015wna,Kromin:2022txz}} in all four loop-momentum components, while our construction analytically integrates the loop energies and subtracts threshold singularities in the real spatial loop-momentum domain.
Our earlier work demonstrated direct numerical integration of the $N_f$ two-loop virtual correction for vector-boson production through $2\to3$~\mbox{\cite{Kermanschah:2024utt}}. 
Subsequent work covered additional fermion-loop topologies with massless and massive internal quarks~\mbox{\cite{Kermanschah:2025wlo,Kermanschah:2026dzb,Kermanschah:2026odl}}.
\par

To study how the complete calculation scales with multiplicity, we develop a new automated threshold subtraction, replacing the manual choice of subtraction projections in our previous implementation~\mbox{\cite{Kermanschah:2024utt}}.
Diagram generation, local UV renormalisation and IR factorisation, threshold subtraction, and numerical integration follow the same rules at every multiplicity, without new integral reductions.
Stochastic sampling of helicities and photon permutations makes the discrete sums manageable at high multiplicity.
We compute the one-loop and two-loop \(N_f\) virtual corrections to the squared matrix elements at fixed phase-space points through \(n=8\).
This reaches a ten-point two-loop calculation with \(1\,733\,760\) bare Feynman diagrams, in addition to the local counterterms.
The same representation yields inclusive and differential hadronic virtual corrections through five photons by direct integration with parton distribution functions (PDFs).
\par
To our knowledge, this is the first automated, multiplicity-independent construction of a gauge-invariant two-loop QCD contribution for arbitrary helicities.
\par

\section{Finite remainders and virtual corrections}
\label{sec:method}

We follow the conventions of Ref.~\mbox{\cite{Kermanschah:2024utt}} and expose charge and colour factors in the UV-renormalised amplitude,
\begin{equation}
  \begin{aligned}
M_{ij,h}
=
&\delta_{ij}(eQ_q)^n
\bigg[
A_h^{(0)}
+ \left(\frac{\alpha_s(\mu_R)}{4\pi}\right) C_F\,A_h^{(1)} \\
&+ \left(\frac{\alpha_s(\mu_R)}{4\pi}\right)^2
C_F T_F N_f\,A_h^{(2,N_f)}
+ \cdots
\bigg].
  \end{aligned}
  \label{eq:amplitude_conventions}
\end{equation}
Here \(N_f\) counts massless quark flavours in the gluon vacuum-polarisation insertion, with all external photons attached to the open quark line.
Other colour and charge structures are omitted.
\par
After universal IR subtraction, let \(H_h^{(i)}\) denote the hard finite remainders and define the respective contributions to the squared matrix element by
\begin{equation}
B
=
\sum_h
\left|A_h^{(0)}\right|^2,
\quad
V^{(i)}
=
\sum_h
2\,{\rm Re}\!\left[
\left(A_h^{(0)}\right)^*H_h^{(i)}
\right],
\label{eq:reduced_coefficients}
\end{equation}
for $i\in\{1,(2,N_f)\}$.
The coefficients $V^{(i)}$ are finite virtual contributions.
Their scheme conventions are specified below.
\par

Using collinear factorisation~\mbox{\cite{Collins:1985ue,Collins:1989gx}}, we define the phase-space integrated PDF convolution by
\begin{equation}
 \langle X\rangle_n
 =
 \frac{2}{4N_c\,n!}
 \sum_q (eQ_q)^{2n}
 \int\!\frac{
   \dd x_1\dd x_2\dd\Phi_n
 }{2x_1x_2 s}
 f_q^{(1)}f_{\bar q}^{(2)}F_nX ,
 \label{eq:hadronic_convolution}
\end{equation}
where \(f_a^{(i)}=f_a(x_i,\mu_F)\) and \(F_n\) is a beam-symmetric measurement function.
The prefactor accounts for spin and colour averaging, the identical-photon symmetry factor, and the reversed beam assignment.
The corresponding Born and finite virtual cross-section contributions are
\begin{equation}
 \begin{aligned}
 \sigma_B&=\langle B\rangle_n,\\
 \delta\sigma^{(1)}&=\left(\frac{\alpha_s(\mu_R)}{4\pi}\right)C_F\langle V^{(1)}\rangle_n,\\
 \delta\sigma^{(2,N_f)}
 &=\left(\frac{\alpha_s(\mu_R)}{4\pi}\right)^2C_FT_FN_f\langle V^{(2,N_f)}\rangle_n .
 \end{aligned}
 \label{eq:virtual_cross_sections}
\end{equation}

We reuse the local IR and UV counterterms and their integrated contributions from Ref.~\mbox{\cite{Kermanschah:2024utt}} and integrate loop energies analytically with the algorithm of Ref.~\mbox{\cite{Kermanschah:2026dzb}}.
The threshold subtraction below makes the integrand locally integrable in four spacetime dimensions, allowing the causal regulator to be removed before numerical integration.
\par

\section{Threshold subtraction}
Let \(\mathcal I(\mathbf k)\) denote the loop-energy-integrated integrand, with spatial loop momenta
\(\mathbf k=(\vec k_1,\ldots,\vec k_\ell)\in\mathbb R^{3\ell}\), where \(\ell=1,2\) is the number of loops.
Its threshold surfaces are denoted by \(t_i(\mathbf k)=0\), with \(t_i<0\) in their interiors.
Following the threshold-subtraction method of Refs.~\mbox{\cite{Kermanschah:2021wbk,Kermanschah:2024utt,Vicini:2024ecf,Kermanschah:2025wlo}}, we parameterise the threshold singularities via radial projections.
We construct the dispersive integrand
\begin{equation}
  \operatorname{Disp}\mathcal I(\mathbf k)
  =\mathcal I(\mathbf k)-\sum_i\sum_{b\in B_i}C_{i,b}(\mathbf k),
  \label{eq:combined_threshold_subtraction}
\end{equation}
where \(B_i\) is the set of radial projections chosen for threshold \(t_i\).
To preserve local cancellations at higher-order pole intersections, participating thresholds must share radial projections~\mbox{\cite{Kermanschah:2021wbk}}.
We generate candidate projections automatically from all oriented bases of the cut-propagator momenta, as higher-order poles, if kinematically allowed, originate from such threshold intersections.
Lower-rank bases are completed to full \(\ell\)-dimensional loop-momentum bases.
Thresholds sharing the same completed, oriented basis use the same basis projection.
Redundant basis projections whose threshold sets are strict subsets of others are discarded.
\par
Each projection \(b\) defines a radial parametrisation of the full \(3\ell\)-dimensional loop-momentum space through its inverse basis map\footnote{A factorised alternative projects only the three-dimensional cut-momentum subspace, keeping the complementary loop momentum fixed and using the \(d=3\) kernel in Eq.~\eqref{eq:inversion_kernel}, but gave approximately \(10\%\) larger fixed-point Monte Carlo errors in our two-loop \(N_f\) tests.}
\begin{equation}
  \begin{aligned}
  \Phi_b(\mathbf k_b)&=S_b^{-1}(\mathbf k_b-\mathbf s_b),
  \qquad
  \mathbf k&=\Phi_b(r_b\mathbf u_b)
  \end{aligned}
  \label{eq:projection_chart}
\end{equation}
where $r_b=|\mathbf k_b|\geq 0$ is the radial coordinate and $\mathbf u_b$ the unit direction in basis \(b\).
The rows of \(S_b\) are the loop-momentum signatures and \(\mathbf s_b\) contains the external shifts.
For the massless cut propagators considered here, the origin \(\Phi_b(\mathbf0)\) lies inside every threshold using basis projection \(b\).
Along a fixed direction \(\mathbf u_b\), the positive on-shell radius $r^*_{i,b}$ is defined via
\begin{equation}
  t_i\!\left(\Phi_b\!\left(r^*_{i,b}(\mathbf u_b)\mathbf u_b\right)\right)=0,
  \qquad r^*_{i,b}(\mathbf u_b)>0.
  \label{eq:projected_radius}
\end{equation}
The radial residue and its counterterm are
\begin{equation}
  \begin{aligned}
  R_{i,b}(\mathbf u_b)
  &=\Res\!\left[\mathcal I\!\left(\Phi_b(r_b\mathbf u_b)\right),
    r_b=r^*_{i,b}(\mathbf u_b)\right],
  \\
  C_{i,b}(\mathbf k)
  &=w_{i,b}\!\left(\Phi_b(r^*_{i,b}\mathbf u_b)\right)
    R_{i,b}(\mathbf u_b)\mathcal K_d(r_b,r^*_{i,b}).
  \end{aligned}
  \label{eq:projection_counterterm}
\end{equation}
In \(C_{i,b}\), \(r_b\) and \(\mathbf u_b\) are understood as functions of \(\mathbf k\).
The kernel
\begin{equation}
  \mathcal K_d(r,r^*)
  =\frac{2^d(r^*)^{2d-1}}
  {\bigl[r^2+(r^*)^2\bigr]^{d-1}\bigl[r^2-(r^*)^2\bigr]}, \quad d=3\ell
  \label{eq:inversion_kernel}
\end{equation}
has unit residue and inherits the causal prescription implicitly through the radial pole \(r^*\).
Its radial principal value vanishes, \(\operatorname{PV}\!\int_0^\infty\!\dd r\,r^{d-1}\mathcal K_d(r,r^*)=0\), by the inversion criterion of Ref.~\mbox{\cite{Kermanschah:2026dzb}}, so the integrated counterterm is the purely absorptive residue.
After interference with the tree amplitude and summation over helicities, the numerator is real.
Thus, the dispersive contribution is real and the absorptive contribution purely imaginary, so the latter drops out of the squared matrix element considered here.
With axial couplings, the absorptive part can contribute to the interference and must then be included~\mbox{\cite{Kermanschah:2024utt}}.
\par

For each \(b\in B_i\), we define weights that distribute the residue of \(t_i\) at common loop momentum
\begin{equation}
    \widetilde w_{i,b}(\mathbf k)
    =
      \sum_{\substack{j\neq i\\b\in B_j}}
      \!\left[\frac{E_{\rm cm}}{t_j(\mathbf k)}\right]^2
    \!,\ \ \
    w_{i,b}(\mathbf k)
    =\frac{\widetilde w_{i,b}(\mathbf k)}
    {\displaystyle\sum_{b'\in B_i}\widetilde w_{i,b'}(\mathbf k)}
  .
  \label{eq:projection_weights}
\end{equation}
The sum includes the other thresholds sharing the basis projection \(b\). If there are none, we set \(\widetilde w_{i,b}=1\).
In each counterterm, \(w_{i,b}\) is evaluated at the projection $b$ of the sampled loop momenta \(\mathbf k\) onto \(t_i=0\).
On \(t_i=0\), all projected points coincide and the weights sum to one, reproducing the full residue.
Away from the surface, the projected points differ and the counterterm weights need not sum to one.
\par

The weights implement the same subtraction principle as the partitions \(\mathcal I=\sum_b m_b\mathcal I\) of Refs.~\mbox{\cite{Kermanschah:2024utt,Kermanschah:2025wlo}}: the on-shell values of \(m_b\) play the role of \(w_{i,b}\), distributing each threshold residue among projections.
Those implementations used manually constructed groups and product weights.
We generate the projections automatically and use the sum in Eq.~\eqref{eq:projection_weights}.
In our tests at high massless multiplicities, product weights sharpened projection selection and amplified counterterm fluctuations, giving larger variances.
\par
Here, only the threshold counterterms are weighted.
We evaluate \(\mathcal I\) once and combine all terms at the same loop momentum, as in Ref.~\mbox{\cite{Kermanschah:2021wbk}}.
Fluctuations from partitioning \(\mathcal I\) cancel pointwise in the sum, but add variance when its weighted pieces are integrated separately.
\par
At isolated intersections containing only additive poles, \(\mathcal I=A/t_i+B/t_j\), the poles can be subtracted independently.
Unlike the construction of Ref.~\mbox{\cite{Kermanschah:2024utt}}, this requires neither a shared projection nor a global partition of \(\mathcal I\).
Shared projections are required by compatible on-shell cuts, whose simultaneous existence is Lorentz invariant, rather than by frame-dependent geometric overlaps.
\par

At an isolated double pole, such as \(1/(t_it_j)\), non-shared projections are quadratically suppressed, and the two residues acquire the same limiting weight \(w_b\) in each shared projection, so
\begin{equation}
  w_{i,b}R_{i,b}+w_{j,b}R_{j,b}
  \longrightarrow
  w_b\left(R_{i,b}+R_{j,b}\right).
\end{equation}
Along the intersection ray, the projected radii and kernels coincide, so equal weights preserve local counterterm cancellation~\mbox{\cite{Kermanschah:2021wbk}}.
\par

For the photon amplitudes considered here, double-pole intersections occur only as collinear tangencies for massless photons.
Massive photons admit geometric overlaps but no double-pole intersections.
This construction applies generally to amplitudes with massless propagators and only one-loop-type thresholds.
Threshold subtraction itself also accommodates genuine multi-loop thresholds and massive propagators~\mbox{\cite{Kermanschah:2025wlo,Vicini:2024ecf}}, although these more general configurations may require extending the automatic construction.
\par

\section{Numerical results}
\label{sec:results}

\begin{table*}[htbp]
 \caption{Fixed-point benchmarks using \(10^9\) samples per integration.
 Results at two additional phase-space points are provided in the ancillary material.}
  \label{tab:fixed_point_benchmarks}
 \centering
 \normalsize
 \setlength{\tabcolsep}{2pt}
 \setlength{\arrayrulewidth}{0.3pt}
 \providecommand{\ttlf}{\ttfamily\fontseries{l}\selectfont}
 {\ttlf
 \sisetup{mode=text,text-rm=\ttlf,group-digits=false}
 \begin{tabular*}{\textwidth}{@{\extracolsep{\fill}}l|S[table-format=1.4]@{}l|S[table-format=+2.4(2)]S[table-format=2.5]S[table-format=1.2]S[table-format=2.3]|S[table-format=+3.3(2)]S[table-format=+3.3]S[table-format=1.2]S[table-format=2.3]}
 \hline
\rule[-1ex]{0pt}{4ex} & \multicolumn{2}{c|}{\multirow{2}{*}{$B\ [\mathrm{GeV}^{4-2n}]$}} & \multicolumn{4}{c|}{$V^{(1)}_{\rm BLHA}/B$} & \multicolumn{4}{c}{$V^{(2,N_f)}_{\overline{\rm MS}}/B$} \\
\rule[-1ex]{0pt}{4ex} & \multicolumn{2}{c|}{} & \multicolumn{1}{c}{this work} & \multicolumn{1}{c}{MG5\_aMC} & \multicolumn{1}{c}{$\Delta\ [\sigma]$} & \multicolumn{1}{c|}{$\Delta\ [\%]$} & \multicolumn{1}{c}{this work} & \multicolumn{1}{c}{Reference} & \multicolumn{1}{c}{$\Delta\ [\sigma]$} & \multicolumn{1}{c}{$\Delta\ [\%]$} \\
 \hline
$2\gamma$ & 2.2063 & e+01 & 3.1505 +- 0.0017 & 3.14733 & 1.88 & 0.053 & -28.035 +- 0.023 & -28.045 & 0.44 & 0.081 \\
$2\gamma^*$ & 2.3362 & e+01 & 3.0474 +- 0.0033 & 3.03627 & 3.38 & 0.11 & -27.014 +- 0.043 & -26.970 & 1.01 & 0.16 \\
$3\gamma$ & 3.9166 & e-03 & 3.1694 +- 0.0043 & 3.16440 & 1.16 & 0.14 & -34.934 +- 0.054 & -34.929 & 0.09 & 0.15 \\
$3\gamma^*$ & 4.0277 & e-03 & 5.294 +- 0.011 & 5.29321 & 0.03 & 0.20 & -46.05 +- 0.18 & {\textemdash} & {\textemdash} & 0.39 \\
$4\gamma$ & 8.3265 & e-06 & 6.192 +- 0.023 & 6.17720 & 0.63 & 0.37 & -61.29 +- 0.26 & {\textemdash} & {\textemdash} & 0.42 \\
$4\gamma^*$ & 2.5948 & e-06 & 9.820 +- 0.020 & 9.80429 & 0.77 & 0.21 & -81.41 +- 0.36 & {\textemdash} & {\textemdash} & 0.45 \\
$5\gamma$ & 2.3631 & e-08 & 8.109 +- 0.062 & 8.13451 & 0.42 & 0.77 & -78.81 +- 0.65 & {\textemdash} & {\textemdash} & 0.83 \\
$5\gamma^*$ & 2.3113 & e-09 & 13.463 +- 0.033 & 13.53350 & 2.12 & 0.25 & -111.01 +- 0.90 & {\textemdash} & {\textemdash} & 0.81 \\
$6\gamma$ & 9.6456 & e-11 & 11.98 +- 0.18 & 11.79234 & 1.09 & 1.5 & -108.9 +- 2.8 & {\textemdash} & {\textemdash} & 2.6 \\
$6\gamma^*$ & 3.5867 & e-12 & 19.665 +- 0.071 & 19.70056 & 0.49 & 0.36 & -159.7 +- 2.1 & {\textemdash} & {\textemdash} & 1.3 \\
$7\gamma$ & 8.8162 & e-14 & 12.13 +- 0.38 & 12.15285 & 0.06 & 3.1 & -114.1 +- 7.0 & {\textemdash} & {\textemdash} & 6.1 \\
$7\gamma^*$ & 3.4204 & e-15 & 23.01 +- 0.18 & 23.07580 & 0.39 & 0.76 & -183.4 +- 4.1 & {\textemdash} & {\textemdash} & 2.2 \\
$8\gamma$ & 2.6848 & e-16 & 12.6 +- 2.3 & {\textemdash} & {\textemdash} & 18 & -188 +- 31 & {\textemdash} & {\textemdash} & 16 \\
$8\gamma^*$ & 4.9446 & e-18 & 25.22 +- 0.22 & {\textemdash} & {\textemdash} & 0.87 & -217.2 +- 5.2 & {\textemdash} & {\textemdash} & 2.4 \\
 \hline
 \end{tabular*}
 }

\end{table*}

We present one-loop and two-loop \(N_f\) finite remainders at fixed phase-space points and the corresponding inclusive and differential hadronic virtual corrections.
The fixed-point results provide validation and scaling benchmarks, rather than intermediate inputs to the hadronic calculation, which integrates phase-space and loop momenta simultaneously.
Our implementation builds on the codebase developed for Refs.~\mbox{\cite{Kermanschah:2024utt,Vicini:2024ecf,Kermanschah:2025wlo,Kermanschah:2026dzb,Kermanschah:2026odl}}.
Symbolic preprocessing uses \textsc{Symbolica}~\mbox{\cite{symbolica}}, while analytic loop-energy integration and expression optimisation use \textsc{FORM}~\mbox{\cite{Vermaseren:2000nd,Kuipers:2013pba}}.
Our numerical integrations use double precision and the adaptive Vegas algorithm~\mbox{\cite{Lepage:1977sw}} implemented in the \textsc{Havana} integrator~\mbox{\cite{havana}}.
The loop integrations use multi-channel importance sampling~\mbox{\cite{Kleiss:1994qy,Ohl:1998jn}} with mappings adapted to soft and collinear regions~\mbox{\cite{Kermanschah:2025wlo}}.
\par

The main text and ancillary material report 100 loop integrations: 84 at fixed phase-space points and 16 for hadronic virtual corrections.
Each is corroborated by at least three additional compatible runs varying random seeds, sampling schedules or integration setups.
\par

The results shown for the two-loop \(N_f\) hard remainders are in the modified minimal subtraction (\(\overline{\rm MS}\)) scheme, where soft and collinear singularities of the UV-renormalised amplitude are factorised as \(M=\mathbf ZH\), with the infrared operator \(\mathbf Z\) of Refs.~\mbox{\cite{Becher:2009cu,Chawdhry:2020for}}.
The finite conversion from our local-subtraction convention follows Ref.~\mbox{\cite{Kermanschah:2024utt}}.
The two-loop \(N_f\) finite-remainder calculation, including conversion to \(\overline{\rm MS}\), requires no \(\mathcal O(\epsilon)\) terms of the one-loop hard amplitude.
For the one-loop fixed-point comparison, we instead use the Binoth Les Houches Accord (BLHA) convention~\mbox{\cite{Binoth:2010xt}} to match \textsc{MadGraph5\_aMC@NLO} (\textsc{MG5\_aMC})~\mbox{\cite{Alwall:2014hca}},
\begin{equation}
  V^{(1)}_{\overline{\rm MS}}
  =V^{(1)}_{\rm BLHA}+\frac{\pi^2}{3}B.
  \label{eq:blha_conversion_new}
\end{equation}

\paragraph*{Fixed phase-space points.}
For each multiplicity, we evaluate three random RAMBO~\mbox{\cite{Kleiss:1985gy}} phase-space points for on-shell and equal-mass off-shell photons with \(m_{\gamma^*}=m_Z\).
These phase-space points are provided in an ancillary text file.
The partonic centre-of-mass energy, renormalisation scale, and UV mass are set to \(1\,\mathrm{TeV}\).
Helicity configurations and photon permutations are sampled stochastically.
We use parity to halve the number of sampled helicity configurations and share adaptive grids among helicities and photon permutations with the same endpoints.
All photon permutations of the conjugate tree amplitude are summed explicitly rather than sampled, which we found to reduce the integration variance.
The one- and two-loop integrations use a conservative adaptation schedule with \(n_{\rm start}=10^7\), \(n_{\rm increase}=10^6\), and \(n_{\rm max}=10^9\).
\par

\begin{figure}[b]
 \centering
 \includegraphics[width=\columnwidth]{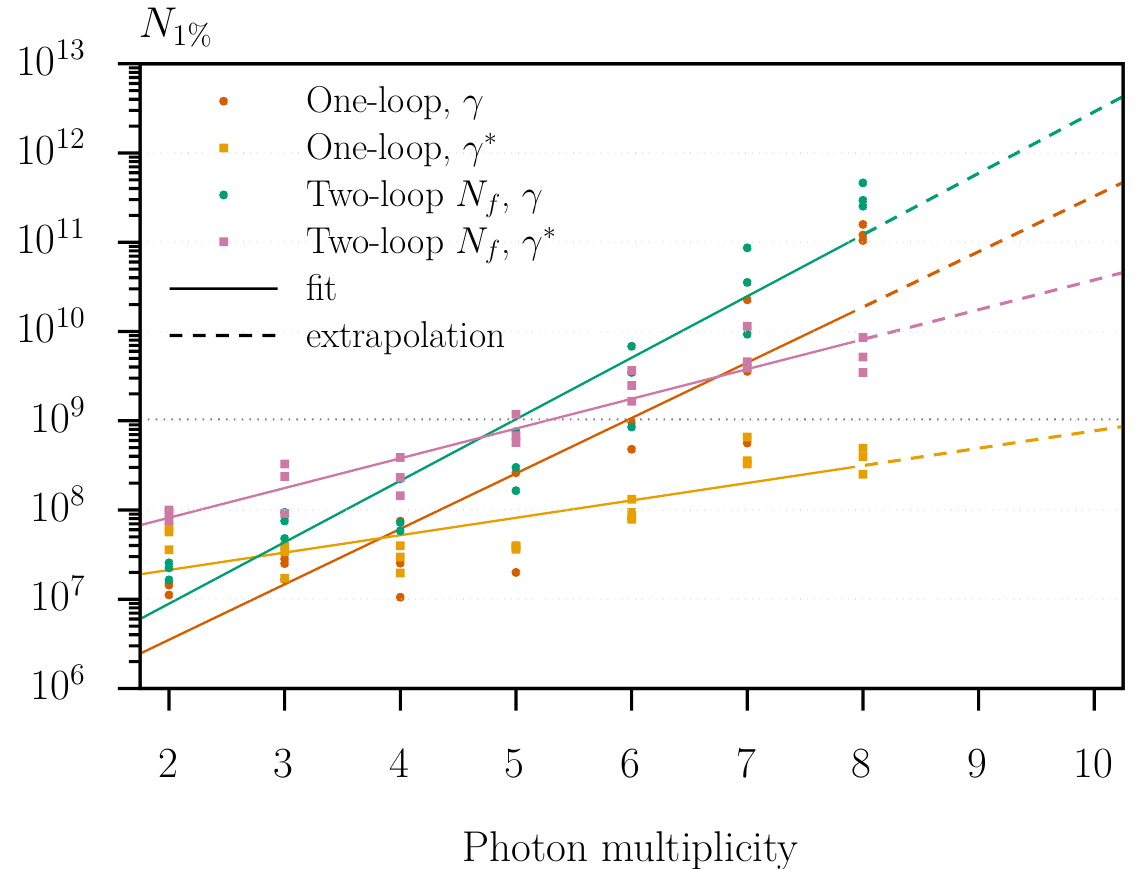}
 \caption{
Sample counts extrapolated to one-percent relative precision for the one-loop and two-loop \(N_f\) finite remainders in our local-subtraction scheme.
Solid curves are log-linear fits over \(n=2,\ldots,8\).
The grey dotted horizontal line marks the \(10^9\) samples used per run.}
 \label{fig:sample_scaling}
\end{figure}

\begin{table}[b]
 \caption{Setup and evaluation costs.
 Generation includes integrand optimisation and compilation includes required dependencies.
 The per-core evaluation times are averaged over 10\,000 calls and quoted to three significant digits.
 The incremental times \(\Delta t_{\rm 1L}\) and \(\Delta t_{\mathrm{2L},N_f}\) subtract the tree evaluation time to estimate fixed-point costs with a cached conjugate tree.
 All timings use 96-core AMD EPYC 9654 processors (maximum 3.71 GHz).
 A full node (192 cores) was allocated for setup.}
  \label{tab:timing_scaling}
 \centering
 \normalsize
 \setlength{\tabcolsep}{2pt}
 \setlength{\arrayrulewidth}{0.3pt}
 \providecommand{\ttlf}{\ttfamily\fontseries{l}\selectfont}
 {\ttlf
 \sisetup{mode=text,text-rm=\ttlf,group-digits=false}
 \begin{tabular*}{\columnwidth}{@{\extracolsep{\fill}}l|S[table-format=4.2]S[table-format=4.1]|S[table-format=2.5]S[table-format=1.5]S[table-format=1.4]}
 \hline
\rule[-1ex]{0pt}{4ex} & \multicolumn{2}{c|}{setup\ [s]} & \multicolumn{3}{c}{evaluation\ [ms]} \\
\rule[-1ex]{0pt}{4ex} & \multicolumn{1}{c}{$t_{\rm gen}$} & \multicolumn{1}{c|}{$t_{\rm comp}$} & \multicolumn{1}{c}{$t_{\rm tree}$} & \multicolumn{1}{c}{$\Delta t_{\rm 1L}$} & \multicolumn{1}{c}{$\Delta t_{\mathrm{2L},N_f}$} \\
 \hline
$2\gamma$ & 3.88 & 75.6 & 0.00175 & 0.00804 & 0.0172 \\
$2\gamma^*$ & 5.46 & 78.4 & 0.00177 & 0.00958 & 0.0197 \\
$3\gamma$ & 5.92 & 79.3 & 0.00272 & 0.0198 & 0.0402 \\
$3\gamma^*$ & 6.76 & 79.7 & 0.00274 & 0.0246 & 0.0486 \\
$4\gamma$ & 10.4 & 85.9 & 0.00676 & 0.0477 & 0.0949 \\
$4\gamma^*$ & 12.0 & 88.8 & 0.00677 & 0.0590 & 0.118 \\
$5\gamma$ & 23.5 & 104 & 0.0297 & 0.116 & 0.222 \\
$5\gamma^*$ & 27.0 & 109 & 0.0298 & 0.129 & 0.263 \\
$6\gamma$ & 68.3 & 138 & 0.190 & 0.241 & 0.500 \\
$6\gamma^*$ & 85.3 & 163 & 0.187 & 0.286 & 0.586 \\
$7\gamma$ & 318 & 310 & 1.49 & 0.490 & 1.09 \\
$7\gamma^*$ & 413 & 443 & 1.43 & 0.606 & 1.26 \\
$8\gamma$ & 2180 & 1640 & 12.6 & 1.26 & 2.75 \\
$8\gamma^*$ & 2820 & 2790 & 12.9 & 1.41 & 2.98 \\
 \hline
 \end{tabular*}
 }

\end{table}

\begin{table*}[t]
 \caption{Inclusive Born cross sections and finite virtual contributions normalised to Born.
 Our results use \(10^{10}\) samples per integration.
 \textsc{MG5\_aMC} results are converted to \(\overline{\rm MS}\).
 Inputs and cuts are specified in the text.}
  \label{tab:inclusive_cross_sections}
 \centering
 \normalsize
 \setlength{\tabcolsep}{1.5pt}
 \setlength{\arrayrulewidth}{0.3pt}
 \providecommand{\ttlf}{\ttfamily\fontseries{l}\selectfont}
 {\ttlf
 \sisetup{mode=text,text-rm=\ttlf,group-digits=false}
 \begin{tabular*}{\textwidth}{@{\extracolsep{\fill}}l|S[table-format=1.6(2)]@{}l|S[table-format=+1.5(2)]S[table-format=+1.5(2)]S[table-format=1.2]S[table-format=2.4]|S[table-format=+1.5(2)]S[table-format=+1.6(2)]S[table-format=1.2]S[table-format=2.4]}
 \hline
\rule[-1ex]{0pt}{4ex} & \multicolumn{2}{c|}{$\sigma_B\ [\mathrm{pb}]$} & \multicolumn{4}{c|}{$\delta\sigma^{(1)}_{\overline{\rm MS}}/\sigma_B\ [\times 10]$} & \multicolumn{4}{c}{$\delta\sigma^{(2,N_f)}_{\overline{\rm MS}}/\sigma_B\ [\times 10^2]$} \\
\rule[-1ex]{0pt}{4ex} & \multicolumn{2}{c|}{} & \multicolumn{1}{c}{this work} & \multicolumn{1}{c}{MG5\_aMC} & \multicolumn{1}{c}{$\Delta\ [\sigma]$} & \multicolumn{1}{c|}{$\Delta\ [\%]$} & \multicolumn{1}{c}{this work} & \multicolumn{1}{c}{Reference} & \multicolumn{1}{c}{$\Delta\ [\sigma]$} & \multicolumn{1}{c}{$\Delta\ [\%]$} \\
 \hline
$2\gamma$ & 2.970663 +- 0.000024 & e+01 & 0.90612 +- 0.00011 & 0.90634 +- 0.00049 & 0.44 & 0.012 & -0.89225 +- 0.00036 & -0.892408 +- 0.000036 & 0.45 & 0.040 \\
$2\gamma^*$ & 2.278761 +- 0.000041 & e+00 & 2.06387 +- 0.00011 & 2.0651 +- 0.0015 & 0.79 & 0.0052 & -1.38914 +- 0.00044 & -1.389203 +- 0.000071 & 0.15 & 0.031 \\
$3\gamma$ & 1.411102 +- 0.000026 & e-02 & 1.56444 +- 0.00044 & 1.5658 +- 0.0011 & 1.12 & 0.028 & -1.4396 +- 0.0016 & -1.43777 +- 0.00026 & 1.11 & 0.11 \\
$3\gamma^*$ & 9.68390 +- 0.00038 & e-04 & 1.61915 +- 0.00047 & 1.6169 +- 0.0011 & 1.84 & 0.029 & -1.2419 +- 0.0033 & {\textemdash} & {\textemdash} & 0.26 \\
$4\gamma$ & 2.313837 +- 0.000073 & e-05 & 1.5755 +- 0.0022 & 1.582 +- 0.013 & 0.50 & 0.14 & -1.6809 +- 0.0086 & {\textemdash} & {\textemdash} & 0.51 \\
$4\gamma^*$ & 8.05407 +- 0.00072 & e-07 & 0.7762 +- 0.0021 & 0.7808 +- 0.0045 & 0.93 & 0.27 & -1.032 +- 0.017 & {\textemdash} & {\textemdash} & 1.7 \\
$5\gamma$ & 4.36430 +- 0.00020 & e-08 & 1.0986 +- 0.0063 & 1.0964 +- 0.0091 & 0.20 & 0.58 & -1.678 +- 0.073 & {\textemdash} & {\textemdash} & 4.4 \\
$5\gamma^*$ & 7.9550 +- 0.0016 & e-10 & -0.217 +- 0.011 & -0.2051 +- 0.0058 & 0.94 & 5.0 & -0.800 +- 0.072 & {\textemdash} & {\textemdash} & 9.0 \\
 \hline
 \end{tabular*}
 }

\end{table*}

Table~\ref{tab:fixed_point_benchmarks} reports the squared Born matrix element and the virtual corrections at a fixed phase-space point.
In the tables, \(n\gamma\) and \(n\gamma^*\) denote \(n\) photons of mass \(0\) and \(m_Z\), respectively.
Parentheses give Monte Carlo errors. In both comparison tables, \(\Delta\ [\sigma]\) gives absolute reference differences in combined-error units and \(\Delta\ [\%]\) our relative Monte Carlo errors.
Where available, \(B\) and \(V^{(1)}_{\rm BLHA}\) are compared with \textsc{MG5\_aMC}.
\textsc{MG5\_aMC} generation and fixed-point evaluation times grew steeply, each reaching about four days at seven photons in our setup.
We therefore did not pursue eight-photon references.
For \(V^{(2,N_f)}_{\overline{\rm MS}}\), we use our implementations of the \(2\gamma\)~\mbox{\cite{Anastasiou:2002zn}} and \(2\gamma^*\)~\mbox{\cite{Gehrmann:2015ora}} amplitudes, with \textsc{lievaluate}~\mbox{\cite{Frellesvig:2016ske}} for the latter.
The massless \(3\gamma\) reference uses \textsc{FivePointAmplitudes}~\mbox{\cite{Abreu:2020cwb,Abreu:2023bdp}} with \textsc{PentagonFunctions++}~\mbox{\cite{Chicherin:2020oor}}.
Results for two more phase-space points are given in the ancillary material.
With a fixed budget of \(10^9\) samples per integration, two-loop uncertainties are around \(1\%\) or smaller through five photons, rising to \(16\!-\!22\%\) for eight on-shell photons and \(1.9\!-\!3.1\%\) for eight virtual photons.
\par

Figure~\ref{fig:sample_scaling} shows the number of samples required for one-percent relative precision, extrapolated assuming \(1/\sqrt N\) scaling.
Table~\ref{tab:timing_scaling} reports the one-time generation and compilation wall times and the evaluation time per sample.
On a single node with 192 allocated CPUs, individual two-loop \(N_f\) fixed-point integrations with approximately \(10^9\) samples take from about ten minutes at \(n=2\) to about six hours at \(n=8\).
\par

\begin{figure}[b]
 \centering
 \includegraphics[width=\columnwidth]{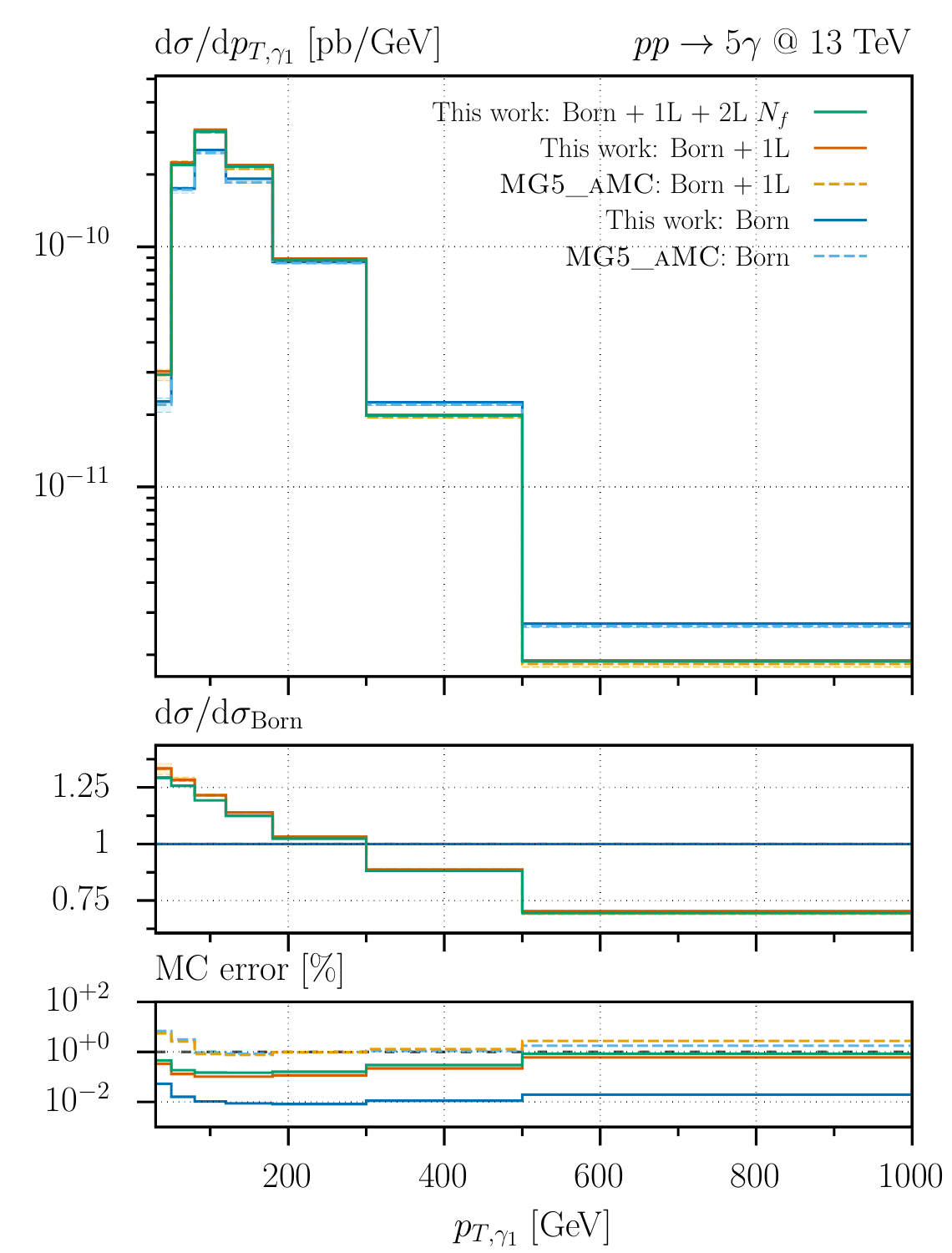}
 \caption{
Leading-photon-\(p_T\) spectrum for the \(q\bar q\)-initiated \(pp\to5\gamma\) contribution.
Curves show Born plus successive one-loop (1L) and two-loop \(N_f\) (\(2\mathrm{L}\,N_f\)) finite virtual contributions in \(\overline{\rm MS}\).
Lower panels show ratios to Born and relative Monte Carlo errors of the cumulative curves.
Corresponding distributions for \(n=2,3,4\) and for \(5\gamma^*\) are provided in the ancillary material.}
 \label{fig:leading_pt_distribution}
\end{figure}

\paragraph*{Phase-space integration.}
Table~\ref{tab:inclusive_cross_sections} shows the inclusive Born cross section and the \(\overline{\rm MS}\) finite virtual contributions normalised to Born for \(n=2,\ldots,5\).
With a fixed budget of \(10^{10}\) samples per integration, inclusive two-loop uncertainties range from \(0.03\%\) to \(9\%\), reaching \(4.4\%\) and \(9.0\%\) for five on-shell and virtual photons, respectively.
\textsc{MG5\_aMC} provides the one-loop references, targeting \(0.1\%\) relative accuracy on the inclusive contributions for \(n=2,3\), relaxed to \(1\%\) for \(n=4,5\) to limit computational cost.
For the two-loop \(N_f\) references, we integrate the same analytic amplitudes with our phase-space integrator.
For the \(q\bar q\)-initiated contributions to \(pp\to n\gamma^{(*)}\), with \(m_{\gamma^*}=m_Z\), we use \(\sqrt{s}=13\,\mathrm{TeV}\), the CT10nlo PDFs (\textsc{LHAPDF} ID 11000)~\mbox{\cite{Lai:2010vv,Buckley:2014ana}} at \(\mu_F=\mu_R=m_Z\), five flavours, \(\alpha_s=0.118\), \(\alpha=1/132.5070\), and \(p_{T,\gamma}>30\,\mathrm{GeV}\).
\par
For our permutation-invariant cuts and observables, all photon orderings in the loop amplitude contribute equally after phase-space integration.
We therefore evaluate one ordering with a factor \(n!\), while retaining the full conjugate tree amplitude.
Whereas the conjugate-tree values are cached at fixed external kinematics, their reevaluation at every sample during phase-space integration dominates the evaluation cost at the highest multiplicities.
We generate the phase space recursively and importance-sample the \(t\)-channel structure and the photon transverse-momentum cut.
The phase-space integrations use \(n_{\rm increase}=10^7\) and \(n_{\rm max}=10^{10}\).
We increase \(n_{\rm start}\) from \(10\) to \(25\), \(50\), \(75\), and \(100\) million samples when the integration dimension exceeds \(4\), \(10\), \(16\), and \(22\), respectively. The analytic references use the same phase-space generator with \(n_{\rm start}=3\times10^6\), \(n_{\rm increase}=3\times10^5\), and \(n_{\rm max}=2\times10^8\).
Including phase-space integration and PDF convolution, \(10^{10}\)-sample two-loop \(N_f\) runs on a node with 192 allocated CPUs take from about three hours at \(n=2\) to around twelve hours at \(n=5\).
\par
Figure~\ref{fig:leading_pt_distribution} shows the leading-photon-\(p_T\) distribution for the massless \(pp\to5\gamma\) contribution, the highest multiplicity for which our uniform setup yields a reliable inclusive two-loop result.
These finite virtual contributions do not constitute complete NLO or NNLO predictions.
With matched PDFs, scales, cuts, binning, and finite-remainder conventions, the Born-normalised one-loop virtual contributions agree with \textsc{MG5\_aMC} through five photons.

\section{Conclusions}
\label{sec:conclusions}
We have automated the calculation of the gauge-invariant fermion-bubble \(N_f\) contribution to the two-loop QCD finite remainder for \(q\bar q\to n\gamma^{(*)}\), from diagram generation to numerical integration.
We compute the corresponding correction to the squared matrix element through ten external legs.
The same representation yields PDF-convolved inclusive and differential virtual corrections, which we demonstrate through five photons.
The calculation combines a new automated, multiplicity-independent threshold-subtraction construction with universal local UV renormalisation and IR factorisation, analytic loop-energy integration, adaptive multi-channel Monte Carlo integration, and stochastic sampling of helicities and photon permutations.
\par
The measured generation, evaluation, and sampling costs show that this contribution can be efficiently generated and integrated within one numerical framework up to high multiplicities.
Increasing the multiplicity requires more computation, but no new integral reductions or master-integral calculations.
\par
The \(N_f\) contribution provides a controlled test of automation at high multiplicity, involving UV subdivergences and overlapping IR singularities, while retaining one-loop-type physical thresholds.
The presented construction extends to \(W\)- and \(Z\)-boson production, with factorisation allowing the electroweak Dirac structures, including \(\gamma_5\), to be evaluated in four dimensions~\mbox{\cite{Kermanschah:2024utt}}.
No \(\mathcal O(\epsilon)\) terms of lower-loop hard amplitudes are required.
Together with known local UV and IR counterterms~\mbox{\cite{Assadsolimani:2009cz,Becker:2010ng}}, these techniques provide the ingredients for automating general one-loop calculations in massless QCD.
The same codebase already treats planar and non-planar contributions from light- and heavy-quark loops with genuine two-loop thresholds~\mbox{\cite{Kermanschah:2025wlo,Kermanschah:2026dzb,Kermanschah:2026odl}}.
Combined with the present high-multiplicity results, this supports the prospect of general two-loop calculations with flexible final-state multiplicity.
Extending this framework to the remaining colour structures and more general threshold configurations is the next decisive step.
\par

\begin{samepage}
\begin{acknowledgments}
I thank Matilde Vicini and Babis Anastasiou for comments on the manuscript.
I am also grateful to Matilde for useful discussions, to Babis for his encouragement and for providing computational resources, and to Fabrizio Caola for supporting my research stay at Oxford.
Computations were performed on the Euler cluster at ETH Zurich.
Codex assisted extensively with research, software development, and writing through complimentary access provided by OpenAI's ChatGPT for Academic Researchers programme.
This work was funded in part by the Swiss National Science Foundation (SNSF) under grant No.~236993.
\end{acknowledgments}
\end{samepage}

\section*{Data availability}
Numerical results and benchmark phase-space points are provided in the article and its ancillary material.

\onecolumngrid
\clearpage
\section*{Ancillary material}
\providecommand{\mainref}[1]{\ref{#1}}

\begin{center}
 \captionof{table}{Fixed-point results at phase-space point 2. Notation and conventions are as in Table~\mainref{tab:fixed_point_benchmarks}.}
  \label{tab:fixed_point_2}
 \centering
 \normalsize
 \setlength{\tabcolsep}{2pt}
 \setlength{\arrayrulewidth}{0.3pt}
 \providecommand{\ttlf}{\ttfamily\fontseries{l}\selectfont}
 {\ttlf
 \sisetup{mode=text,text-rm=\ttlf,group-digits=false}
 \begin{tabular*}{\textwidth}{@{\extracolsep{\fill}}l|S[table-format=1.4]@{}l|S[table-format=+2.4(2)]S[table-format=2.5]S[table-format=1.2]S[table-format=2.3]|S[table-format=+3.3(2)]S[table-format=+3.3]S[table-format=1.2]S[table-format=2.3]}
 \hline
\rule[-1ex]{0pt}{4ex} & \multicolumn{2}{c|}{\multirow{2}{*}{$B\ [\mathrm{GeV}^{4-2n}]$}} & \multicolumn{4}{c|}{$V^{(1)}_{\rm BLHA}/B$} & \multicolumn{4}{c}{$V^{(2,N_f)}_{\overline{\rm MS}}/B$} \\
\rule[-1ex]{0pt}{4ex} & \multicolumn{2}{c|}{} & \multicolumn{1}{c}{this work} & \multicolumn{1}{c}{MG5\_aMC} & \multicolumn{1}{c}{$\Delta\ [\sigma]$} & \multicolumn{1}{c|}{$\Delta\ [\%]$} & \multicolumn{1}{c}{this work} & \multicolumn{1}{c}{Reference} & \multicolumn{1}{c}{$\Delta\ [\sigma]$} & \multicolumn{1}{c}{$\Delta\ [\%]$} \\
 \hline
$2\gamma$ & 1.8454 & e+01 & 3.3658 +- 0.0017 & 3.36604 & 0.15 & 0.050 & -29.762 +- 0.022 & -29.777 & 0.68 & 0.073 \\
$2\gamma^*$ & 1.9633 & e+01 & 3.2150 +- 0.0034 & 3.21172 & 0.94 & 0.11 & -28.565 +- 0.050 & -28.589 & 0.47 & 0.17 \\
$3\gamma$ & 2.6286 & e-02 & 3.3288 +- 0.0059 & 3.32072 & 1.37 & 0.18 & -34.572 +- 0.074 & -34.554 & 0.23 & 0.22 \\
$3\gamma^*$ & 1.6447 & e-02 & 6.4115 +- 0.0086 & 6.40915 & 0.28 & 0.13 & -51.92 +- 0.13 & {\textemdash} & {\textemdash} & 0.25 \\
$4\gamma$ & 2.9867 & e-05 & 14.215 +- 0.017 & 14.23077 & 0.93 & 0.12 & -122.25 +- 0.28 & {\textemdash} & {\textemdash} & 0.23 \\
$4\gamma^*$ & 2.1046 & e-05 & 22.141 +- 0.034 & 22.11399 & 0.80 & 0.15 & -178.02 +- 0.67 & {\textemdash} & {\textemdash} & 0.38 \\
$5\gamma$ & 3.1611 & e-08 & 16.935 +- 0.041 & 16.93631 & 0.02 & 0.24 & -146.78 +- 0.80 & {\textemdash} & {\textemdash} & 0.54 \\
$5\gamma^*$ & 1.4737 & e-08 & 26.546 +- 0.059 & 26.60212 & 0.94 & 0.22 & -214.5 +- 1.6 & {\textemdash} & {\textemdash} & 0.76 \\
$6\gamma$ & 4.4047 & e-12 & 9.60 +- 0.11 & 9.51211 & 0.81 & 1.1 & -91.9 +- 1.7 & {\textemdash} & {\textemdash} & 1.9 \\
$6\gamma^*$ & 1.1814 & e-12 & 17.411 +- 0.084 & 17.51532 & 1.23 & 0.48 & -140.0 +- 2.7 & {\textemdash} & {\textemdash} & 1.9 \\
$7\gamma$ & 1.1212 & e-14 & 9.86 +- 0.85 & 10.41307 & 0.65 & 8.6 & -119 +- 11 & {\textemdash} & {\textemdash} & 9.4 \\
$7\gamma^*$ & 9.5821 & e-16 & 20.71 +- 0.23 & 20.76756 & 0.26 & 1.1 & -163.6 +- 5.7 & {\textemdash} & {\textemdash} & 3.5 \\
$8\gamma$ & 3.9092 & e-17 & 14.9 +- 2.7 & {\textemdash} & {\textemdash} & 18 & -164 +- 29 & {\textemdash} & {\textemdash} & 18 \\
$8\gamma^*$ & 2.6997 & e-18 & 25.35 +- 0.25 & {\textemdash} & {\textemdash} & 0.97 & -204.0 +- 6.3 & {\textemdash} & {\textemdash} & 3.1 \\
 \hline
 \end{tabular*}
 }

 \vspace{1.5em}

 \captionof{table}{Fixed-point results at phase-space point 3. Notation and conventions are as in Table~\mainref{tab:fixed_point_benchmarks}.}
  \label{tab:fixed_point_3}
 \centering
 \normalsize
 \setlength{\tabcolsep}{2pt}
 \setlength{\arrayrulewidth}{0.3pt}
 \providecommand{\ttlf}{\ttfamily\fontseries{l}\selectfont}
 {\ttlf
 \sisetup{mode=text,text-rm=\ttlf,group-digits=false}
 \begin{tabular*}{\textwidth}{@{\extracolsep{\fill}}l|S[table-format=1.4]@{}l|S[table-format=+2.4(2)]S[table-format=2.5]S[table-format=1.2]S[table-format=2.3]|S[table-format=+3.3(2)]S[table-format=+3.3]S[table-format=1.2]S[table-format=2.3]}
 \hline
\rule[-1ex]{0pt}{4ex} & \multicolumn{2}{c|}{\multirow{2}{*}{$B\ [\mathrm{GeV}^{4-2n}]$}} & \multicolumn{4}{c|}{$V^{(1)}_{\rm BLHA}/B$} & \multicolumn{4}{c}{$V^{(2,N_f)}_{\overline{\rm MS}}/B$} \\
\rule[-1ex]{0pt}{4ex} & \multicolumn{2}{c|}{} & \multicolumn{1}{c}{this work} & \multicolumn{1}{c}{MG5\_aMC} & \multicolumn{1}{c}{$\Delta\ [\sigma]$} & \multicolumn{1}{c|}{$\Delta\ [\%]$} & \multicolumn{1}{c}{this work} & \multicolumn{1}{c}{Reference} & \multicolumn{1}{c}{$\Delta\ [\sigma]$} & \multicolumn{1}{c}{$\Delta\ [\%]$} \\
 \hline
$2\gamma$ & 4.2581 & e+01 & 3.1593 +- 0.0017 & 3.15700 & 1.36 & 0.054 & -27.570 +- 0.024 & -27.565 & 0.23 & 0.085 \\
$2\gamma^*$ & 4.4559 & e+01 & 3.2123 +- 0.0027 & 3.21096 & 0.49 & 0.085 & -26.899 +- 0.039 & -26.850 & 1.26 & 0.14 \\
$3\gamma$ & 3.8137 & e-03 & 2.4468 +- 0.0042 & 2.44646 & 0.09 & 0.17 & -30.428 +- 0.054 & -30.424 & 0.09 & 0.18 \\
$3\gamma^*$ & 4.1521 & e-03 & 4.9966 +- 0.0095 & 4.99670 & 0.01 & 0.19 & -43.22 +- 0.19 & {\textemdash} & {\textemdash} & 0.44 \\
$4\gamma$ & 4.6418 & e-06 & 10.554 +- 0.020 & 10.54131 & 0.61 & 0.19 & -88.29 +- 0.22 & {\textemdash} & {\textemdash} & 0.25 \\
$4\gamma^*$ & 2.5784 & e-06 & 12.474 +- 0.029 & 12.48948 & 0.54 & 0.23 & -99.98 +- 0.59 & {\textemdash} & {\textemdash} & 0.59 \\
$5\gamma$ & 3.1836 & e-10 & 6.951 +- 0.016 & 6.96589 & 0.92 & 0.22 & -70.87 +- 0.27 & {\textemdash} & {\textemdash} & 0.38 \\
$5\gamma^*$ & 4.2834 & e-10 & 12.532 +- 0.033 & 12.51324 & 0.58 & 0.26 & -101.9 +- 1.1 & {\textemdash} & {\textemdash} & 1.1 \\
$6\gamma$ & 1.2919 & e-11 & 11.979 +- 0.050 & 11.86699 & 2.21 & 0.42 & -105.71 +- 0.98 & {\textemdash} & {\textemdash} & 0.92 \\
$6\gamma^*$ & 4.2281 & e-12 & 21.917 +- 0.085 & 22.01444 & 1.15 & 0.39 & -174.2 +- 2.8 & {\textemdash} & {\textemdash} & 1.6 \\
$7\gamma$ & 1.9354 & e-13 & 17.50 +- 0.19 & 17.43974 & 0.34 & 1.1 & -152.4 +- 4.8 & {\textemdash} & {\textemdash} & 3.1 \\
$7\gamma^*$ & 9.7865 & e-15 & 27.30 +- 0.21 & 27.66318 & 1.75 & 0.76 & -218.8 +- 4.5 & {\textemdash} & {\textemdash} & 2.0 \\
$8\gamma$ & 3.6862 & e-15 & 6.9 +- 2.1 & {\textemdash} & {\textemdash} & 31 & -123 +- 27 & {\textemdash} & {\textemdash} & 22 \\
$8\gamma^*$ & 3.5581 & e-17 & 34.05 +- 0.22 & {\textemdash} & {\textemdash} & 0.64 & -282.6 +- 5.5 & {\textemdash} & {\textemdash} & 1.9 \\
 \hline
 \end{tabular*}
 }

\end{center}

\clearpage

\begin{figure}[p]
 \centering
 \includegraphics[width=.46\textwidth]{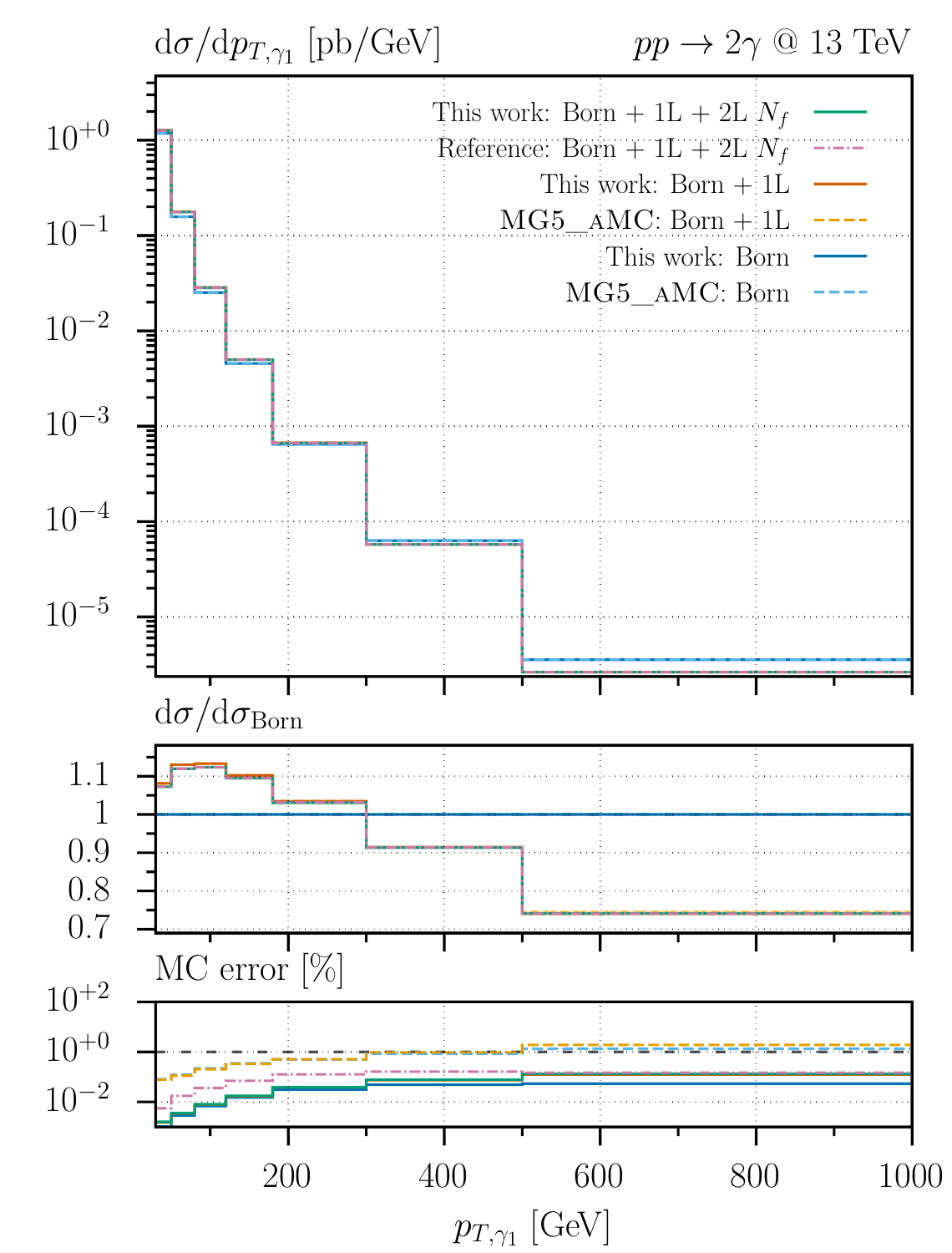}
 \hfill
 \includegraphics[width=.46\textwidth]{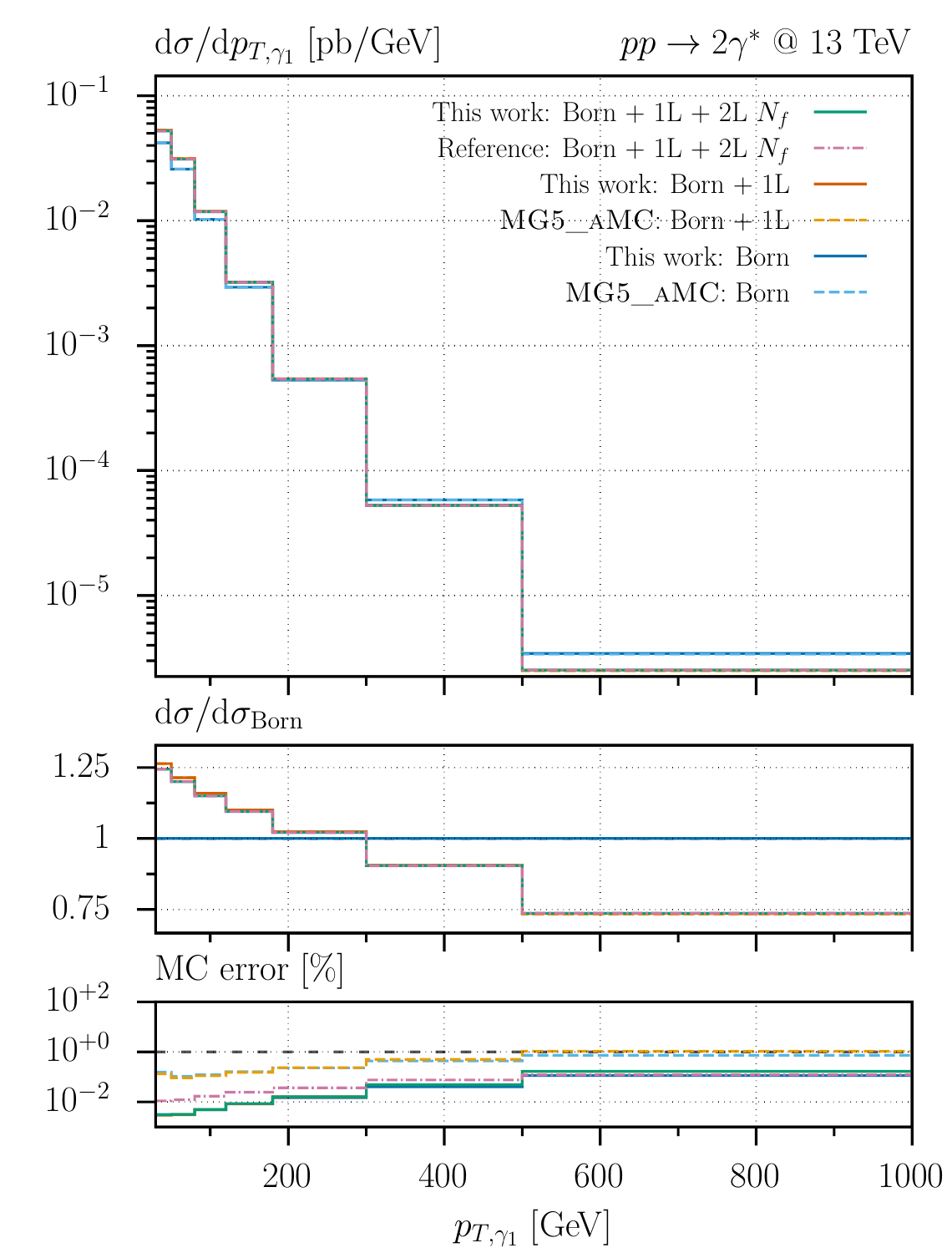}
 \par
 \includegraphics[width=.46\textwidth]{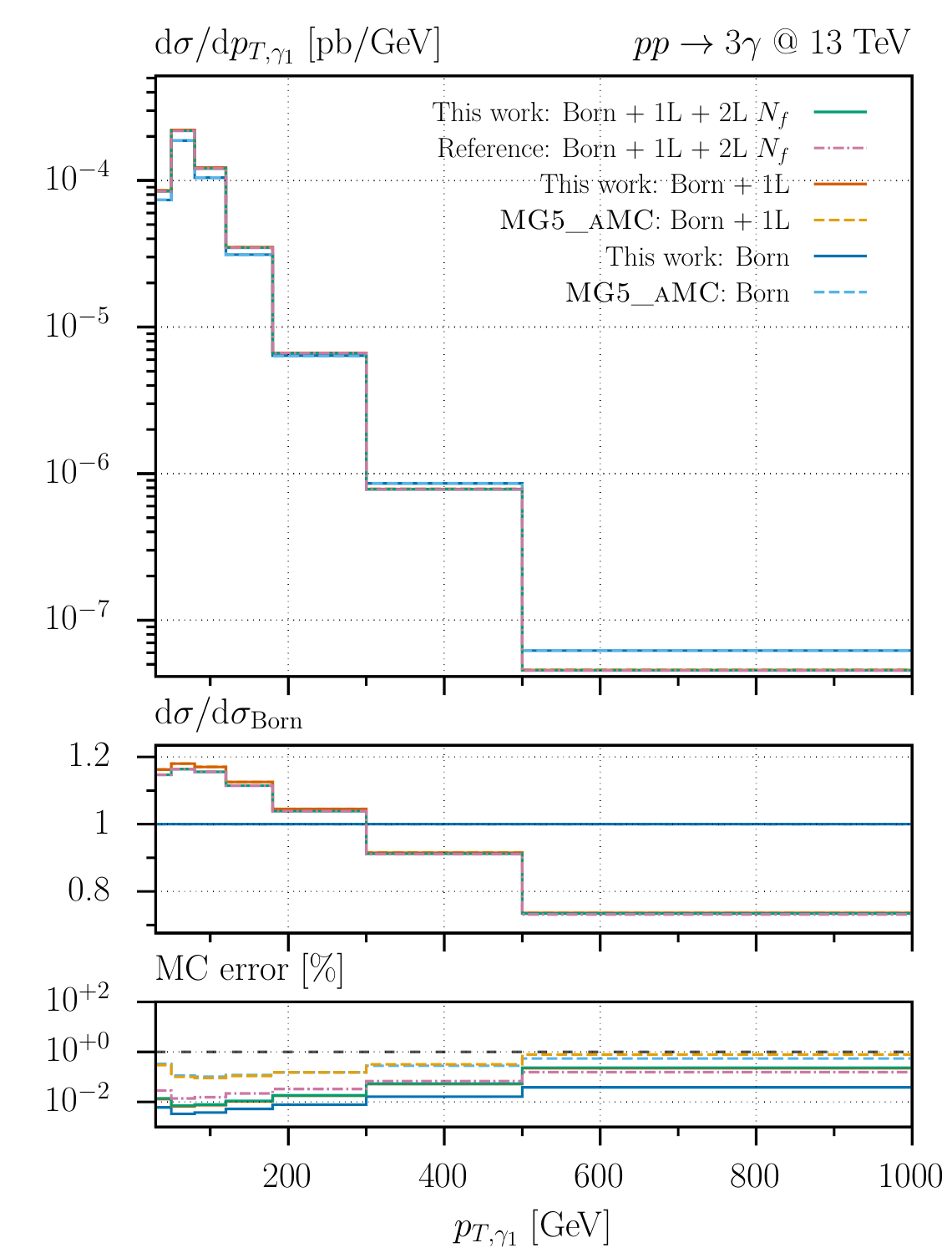}
 \hfill
 \includegraphics[width=.46\textwidth]{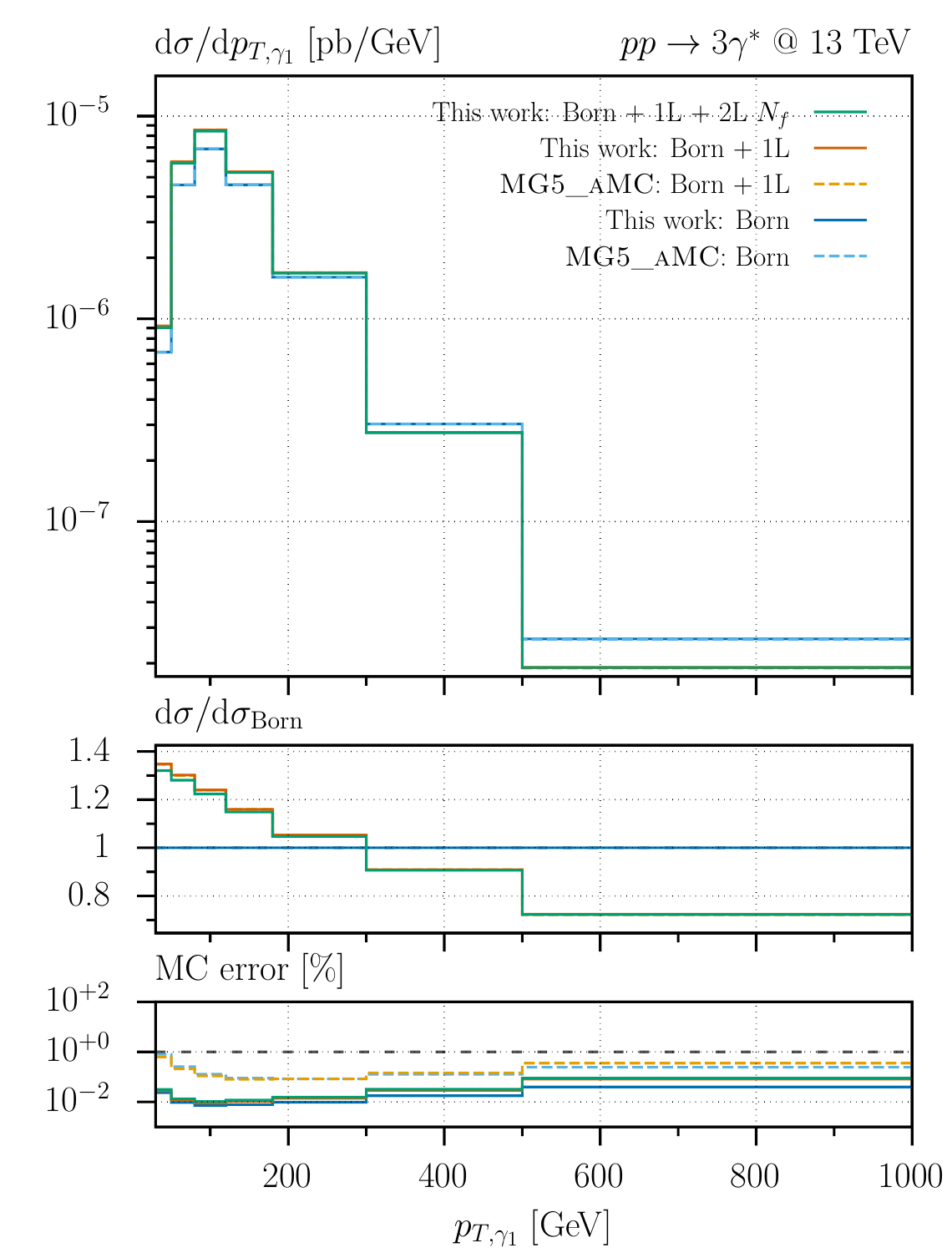}
 \par
 \caption{Combined leading-photon-\(p_T\) distributions for \(n=2,3\). Conventions are as in Fig.~\mainref{fig:leading_pt_distribution}.}
 \label{fig:ancillary_pt_2_3}
\end{figure}

\begin{figure}[p]
 \centering
 \includegraphics[width=.46\textwidth]{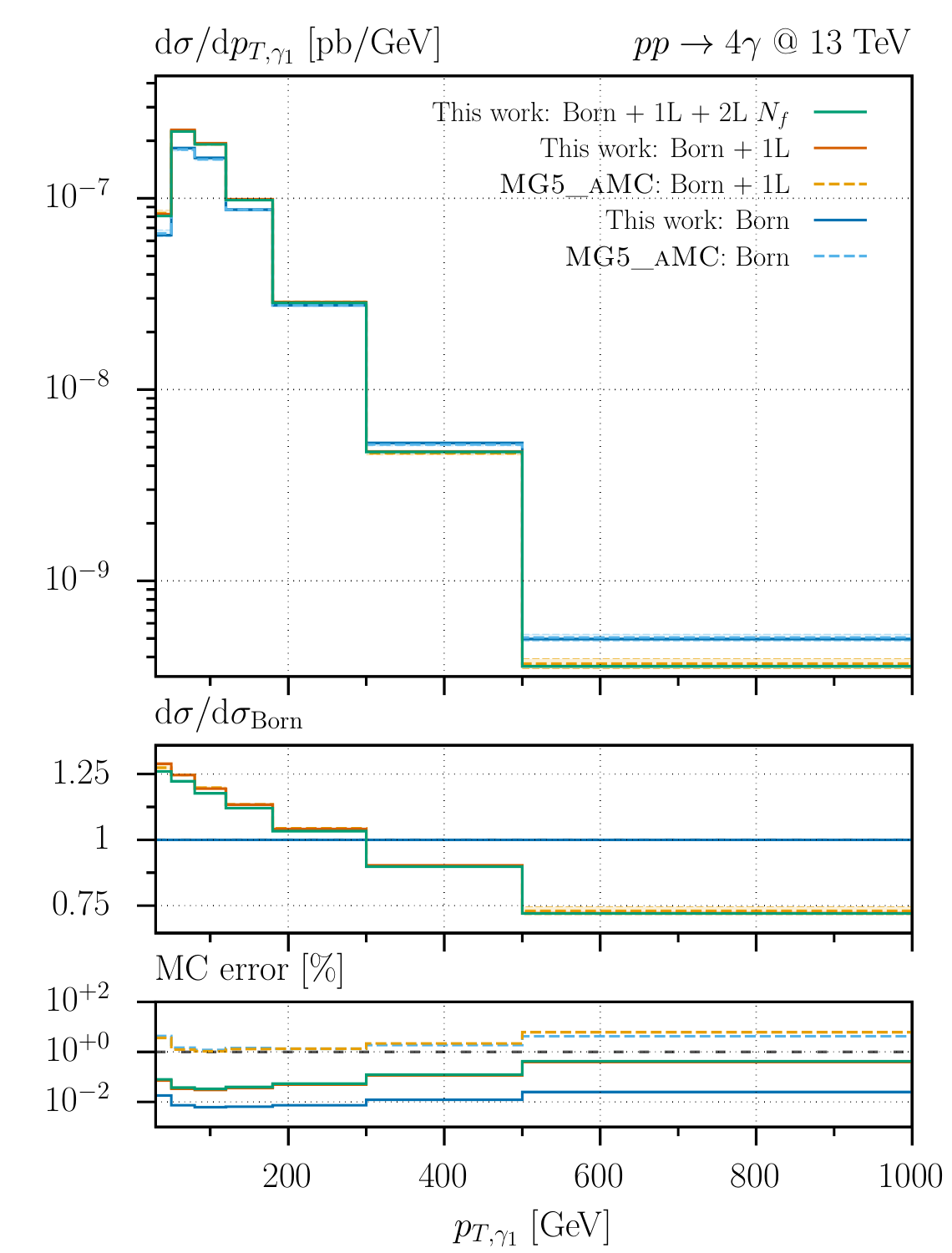}
 \hfill
 \includegraphics[width=.46\textwidth]{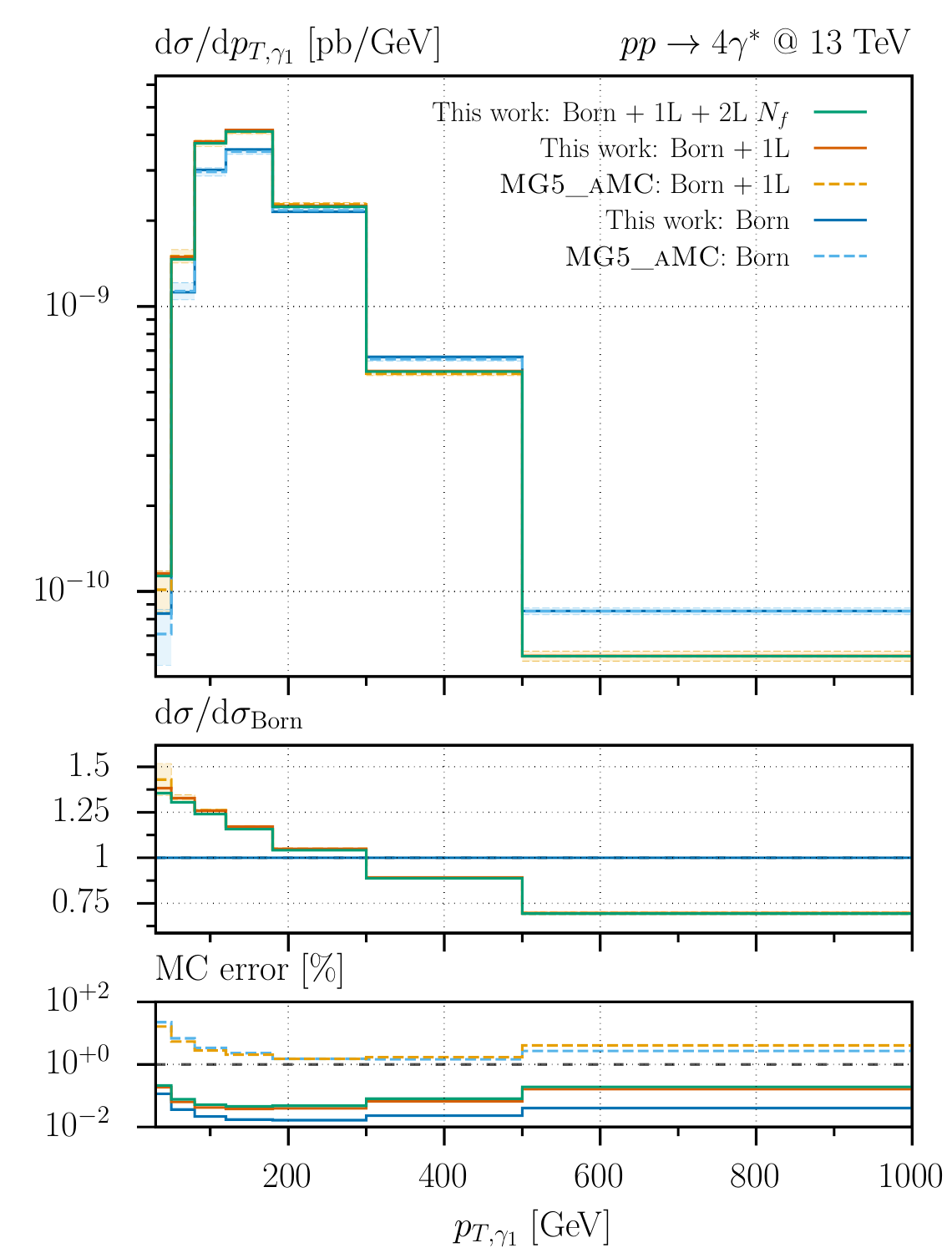}
 \par
 \includegraphics[width=.46\textwidth]{figures/ee_aaaaa_massless_leading_pt_virtual_orders_msbar_rebinned.pdf}\hfill
 \includegraphics[width=.46\textwidth]{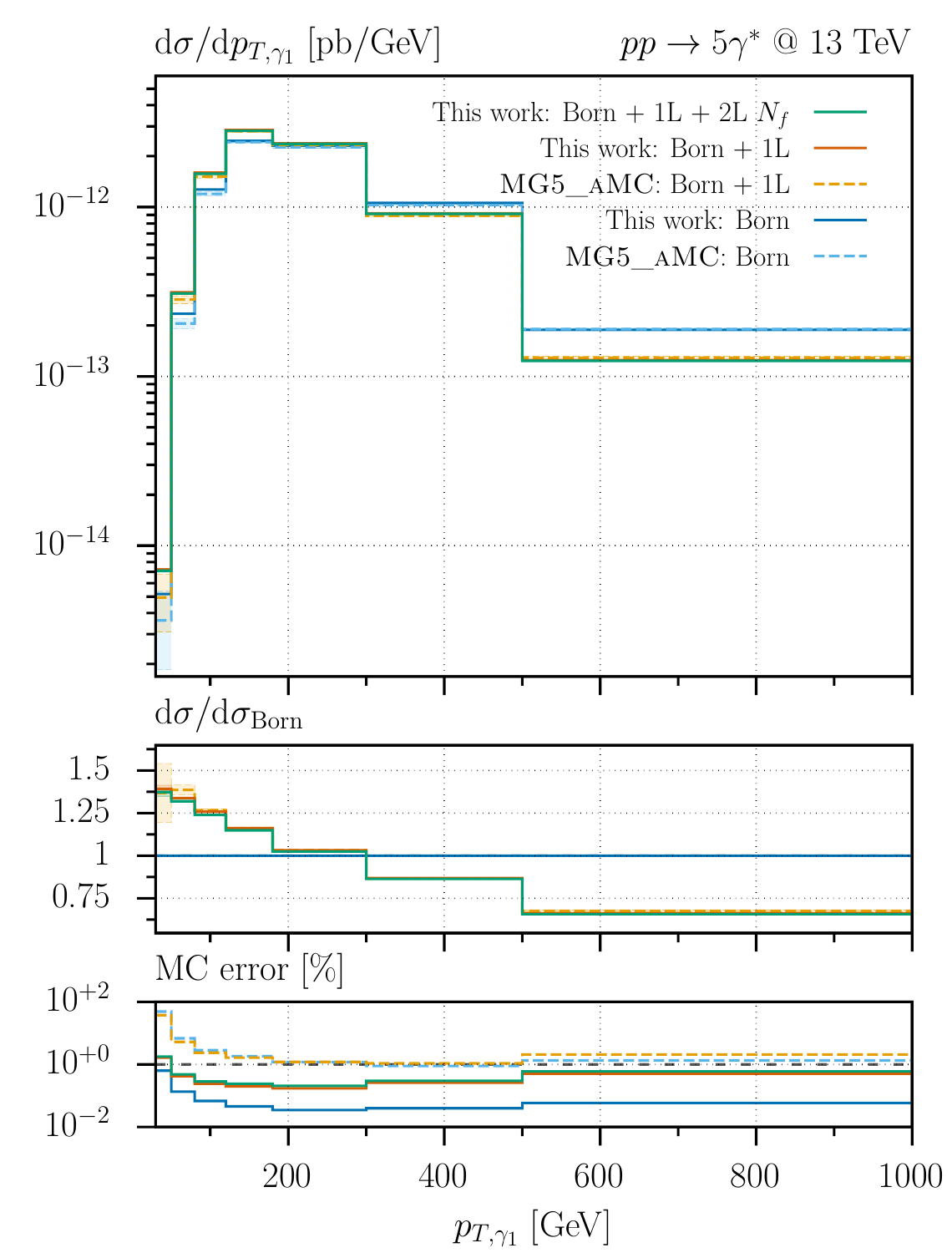}
 \caption{
Combined leading-photon-\(p_T\) distributions for \(n=4,5\).
Conventions are as in Fig.~\mainref{fig:leading_pt_distribution}.}
 \label{fig:ancillary_pt_4_5}
\end{figure}

\clearpage

\twocolumngrid

\bibliography{references}

@article{Kermanschah:2024utt,
  author = {Kermanschah, Dario and Vicini, Matilde},
  title = {{$N_f$-contribution to the virtual correction for electroweak vector boson production at NNLO}},
  eprint = {2407.18051},
  archivePrefix = {arXiv},
  primaryClass = {hep-ph},
  doi = {10.1007/JHEP09(2025)213},
  journal = {JHEP},
  volume = {2025},
  number = {09},
  pages = {213},
  year = {2025}
}

@article{Kermanschah:2025wlo,
  author = {Kermanschah, Dario and Vicini, Matilde},
  title = {{Two-loop QCD corrections for real and off-shell diphoton and triphoton production via quark loops}},
  eprint = {2510.18801},
  archivePrefix = {arXiv},
  primaryClass = {hep-ph},
  doi = {10.1007/JHEP03(2026)004},
  journal = {JHEP},
  volume = {2026},
  number = {03},
  pages = {004},
  year = {2026}
}

@misc{Kermanschah:2026dzb,
  author = {Kermanschah, Dario and Vicini, Matilde},
  title = {{Heavy-quark box-loop corrections to $q\bar q\to Z\gamma$ at two loops in QCD}},
  eprint = {2603.03169},
  archivePrefix = {arXiv},
  primaryClass = {hep-ph},
  year = {2026}
}

@misc{Kermanschah:2026odl,
  author = {Kermanschah, Dario and Vicini, Matilde},
  title = {{Axial triangles in $q\bar q\to Z\gamma$ at two loops in QCD directly in four dimensions}},
  eprint = {2603.03171},
  archivePrefix = {arXiv},
  primaryClass = {hep-ph},
  year = {2026}
}

@article{Becker:2011vg,
  author = {Becker, Sebastian and Goetz, Daniel and Reuschle, Christian and Schwan, Christopher and Weinzierl, Stefan},
  title = {{NLO results for five, six and seven jets in electron-positron annihilation}},
  eprint = {1111.1733},
  archivePrefix = {arXiv},
  primaryClass = {hep-ph},
  doi = {10.1103/PhysRevLett.108.032005},
  journal = {Phys. Rev. Lett.},
  volume = {108},
  pages = {032005},
  year = {2012}
}

@article{Alwall:2014hca,
  author = {Alwall, J. and Frederix, R. and Frixione, S. and Hirschi, V. and Maltoni, F. and Mattelaer, O. and Shao, H.-S. and Stelzer, T. and Torrielli, P. and Zaro, M.},
  title = {{The automated computation of tree-level and next-to-leading order differential cross sections, and their matching to parton shower simulations}},
  eprint = {1405.0301},
  archivePrefix = {arXiv},
  primaryClass = {hep-ph},
  doi = {10.1007/JHEP07(2014)079},
  journal = {JHEP},
  volume = {2014},
  number = {07},
  pages = {079},
  year = {2014}
}

@article{Capatti:2019ypt,
  author = {Capatti, Zeno and Hirschi, Valentin and Kermanschah, Dario and Ruijl, Ben},
  title = {{Loop-Tree Duality for Multiloop Numerical Integration}},
  eprint = {1906.06138},
  archivePrefix = {arXiv},
  primaryClass = {hep-ph},
  doi = {10.1103/PhysRevLett.123.151602},
  journal = {Phys. Rev. Lett.},
  volume = {123},
  pages = {151602},
  year = {2019}
}

@article{Kermanschah:2021wbk,
  author = {Kermanschah, Dario},
  title = {{Numerical integration of loop integrals through local cancellation of threshold singularities}},
  eprint = {2110.06869},
  archivePrefix = {arXiv},
  primaryClass = {hep-ph},
  doi = {10.1007/JHEP01(2022)151},
  journal = {JHEP},
  volume = {2022},
  number = {01},
  pages = {151},
  year = {2022}
}

@article{Anastasiou:2002zn,
    author = "Anastasiou, C. and Glover, E. W. Nigel and Tejeda-Yeomans, M. E.",
    title = "{Two loop QED and QCD corrections to massless fermion boson scattering}",
    eprint = "hep-ph/0201274",
    archivePrefix = "arXiv",
    reportNumber = "SLAC-PUB-9130, DCPT-02-10, IPPP-02-05, YITP-SB-02-04",
    doi = "10.1016/S0550-3213(02)00140-2",
    journal = "Nucl. Phys. B",
    volume = "629",
    pages = "255--289",
    year = "2002"
}

@article{Gehrmann:2015ora,
    author = "Gehrmann, Thomas and von Manteuffel, Andreas and Tancredi, Lorenzo",
    title = "{The two-loop helicity amplitudes for $q\overline{q}^{\prime}\to V_1 V_2\to 4$ leptons}",
    eprint = "1503.04812",
    archivePrefix = "arXiv",
    primaryClass = "hep-ph",
    reportNumber = "ZU-TH-03-15, MITP-15-011, TTP15-011",
    doi = "10.1007/JHEP09(2015)128",
    journal = "JHEP",
    volume = "2015",
    number = "09",
    pages = "128",
    year = "2015"
}

@article{Frellesvig:2016ske,
    author = "Frellesvig, Hjalte and Tommasini, Damiano and Wever, Christopher",
    title = "{On the reduction of generalized polylogarithms to $\mathrm{Li}_n$ and $\mathrm{Li}_{2,2}$ and on the evaluation thereof}",
    eprint = "1601.02649",
    archivePrefix = "arXiv",
    primaryClass = "hep-ph",
    doi = "10.1007/JHEP03(2016)189",
    journal = "JHEP",
    volume = "2016",
    number = "03",
    pages = "189",
    year = "2016"
}

@article{Chicherin:2020oor,
    author = "Chicherin, Dmitry and Sotnikov, Vasily",
    title = "{Pentagon Functions for Scattering of Five Massless Particles}",
    eprint = "2009.07803",
    archivePrefix = "arXiv",
    primaryClass = "hep-ph",
    reportNumber = "MPP-2020-171",
    doi = "10.1007/JHEP12(2020)167",
    journal = "JHEP",
    volume = "2020",
    number = "12",
    pages = "167",
    year = "2020"
}

@article{Abreu:2020cwb,
    author = "Abreu, S. and Page, B. and Pascual, E. and Sotnikov, V.",
    title = "{Leading-Color Two-Loop QCD Corrections for Three-Photon Production at Hadron Colliders}",
    eprint = "2010.15834",
    archivePrefix = "arXiv",
    primaryClass = "hep-ph",
    reportNumber = "CERN-TH-2020-176, FR-PHENO-2020-016, MPP-2020-190",
    doi = "10.1007/JHEP01(2021)078",
    journal = "JHEP",
    volume = "2021",
    number = "01",
    pages = "078",
    year = "2021"
}

@article{Abreu:2023bdp,
    author = "Abreu, Samuel and De Laurentis, Giuseppe and Ita, Harald and Klinkert, Maximillian and Page, Ben and Sotnikov, Vasily",
    title = "{Two-loop QCD corrections for three-photon production at hadron colliders}",
    eprint = "2305.17056",
    archivePrefix = "arXiv",
    primaryClass = "hep-ph",
    reportNumber = "CERN-TH-2023-090, PSI-PR-23-16, ZU-TH 23/23",
    doi = "10.21468/SciPostPhys.15.4.157",
    journal = "SciPost Phys.",
    volume = "15",
    number = "4",
    pages = "157",
    year = "2023"
}

@article{Bogoliubov:1957gp,
    author = "Bogoliubov, N. N. and Parasiuk, O. S.",
    title = "{On the Multiplication of the causal function in the quantum theory of fields}",
    doi = "10.1007/BF02392399",
    journal = "Acta Math.",
    volume = "97",
    pages = "227--266",
    year = "1957"
}

@article{Hepp:1966eg,
    author = "Hepp, Klaus",
    title = "{Proof of the Bogolyubov-Parasiuk theorem on renormalization}",
    doi = "10.1007/BF01773358",
    journal = "Commun. Math. Phys.",
    volume = "2",
    pages = "301--326",
    year = "1966"
}

@article{Zimmermann:1969jj,
    author = "Zimmermann, W.",
    title = "{Convergence of Bogolyubov's method of renormalization in momentum space}",
    doi = "10.1007/BF01645676",
    journal = "Commun. Math. Phys.",
    volume = "15",
    pages = "208--234",
    year = "1969"
}

@article{Anastasiou:2020sdt,
    author = "Anastasiou, Charalampos and Haindl, Rayan and Sterman, George and Yang, Zhou and Zeng, Mao",
    title = "{Locally finite two-loop amplitudes for off-shell multi-photon production in electron-positron annihilation}",
    eprint = "2008.12293",
    archivePrefix = "arXiv",
    primaryClass = "hep-ph",
    reportNumber = "YITP-SB-2020-25",
    doi = "10.1007/JHEP04(2021)222",
    journal = "JHEP",
    volume = "2021",
    number = "04",
    pages = "222",
    year = "2021"
}

@article{Anastasiou:2022eym,
    author = "Anastasiou, Charalampos and Sterman, George",
    title = "{Locally finite two-loop QCD amplitudes from IR universality for electroweak production}",
    eprint = "2212.12162",
    archivePrefix = "arXiv",
    primaryClass = "hep-ph",
    reportNumber = "YITP-SB-2022-43",
    doi = "10.1007/JHEP05(2023)242",
    journal = "JHEP",
    volume = "2023",
    number = "05",
    pages = "242",
    year = "2023"
}

@article{Assadsolimani:2009cz,
    author = "Assadsolimani, Mohammad and Becker, Sebastian and Weinzierl, Stefan",
    title = "{A Simple formula for the infrared singular part of the integrand of one-loop QCD amplitudes}",
    eprint = "0912.1680",
    archivePrefix = "arXiv",
    primaryClass = "hep-ph",
    reportNumber = "MZ-TH-09-36",
    doi = "10.1103/PhysRevD.81.094002",
    journal = "Phys. Rev. D",
    volume = "81",
    pages = "094002",
    year = "2010"
}

@article{Catani:2008xa,
    author = "Catani, Stefano and Gleisberg, Tanju and Krauss, Frank and Rodrigo, German and Winter, Jan-Christopher",
    title = "{From loops to trees by-passing Feynman's theorem}",
    eprint = "0804.3170",
    archivePrefix = "arXiv",
    primaryClass = "hep-ph",
    reportNumber = "FERMILAB-PUB-08-092-T, IFIC-08-21, IPPP-08-22, SLAC-PUB-13218",
    doi = "10.1088/1126-6708/2008/09/065",
    journal = "JHEP",
    volume = "2008",
    number = "09",
    pages = "065",
    year = "2008"
}

@article{Capatti:2022mly,
    author = "Capatti, Zeno",
    title = "{Exposing the threshold structure of loop integrals}",
    eprint = "2211.09653",
    archivePrefix = "arXiv",
    primaryClass = "hep-th",
    doi = "10.1103/PhysRevD.107.L051902",
    journal = "Phys. Rev. D",
    volume = "107",
    number = "5",
    pages = "L051902",
    year = "2023"
}

@article{Sterman:2023xdj,
    author = "Sterman, George and Venkata, Aniruddha",
    title = "{Local infrared safety in time-ordered perturbation theory}",
    eprint = "2309.13023",
    archivePrefix = "arXiv",
    primaryClass = "hep-ph",
    reportNumber = "YITP-SB-2023-28",
    doi = "10.1007/JHEP02(2024)101",
    journal = "JHEP",
    volume = "2024",
    number = "02",
    pages = "101",
    year = "2024"
}

@article{Vicini:2024ecf,
    author = "Vicini, Matilde and Kermanschah, Dario",
    title = "{Numerical integration of the double- and triple-box integrals using threshold subtraction}",
    eprint = "2407.21511",
    archivePrefix = "arXiv",
    primaryClass = "hep-ph",
    doi = "10.22323/1.467.0078",
    journal = "PoS",
    volume = "LL2024",
    pages = "078",
    year = "2024"
}

@article{Lepage:1977sw,
    author = "Lepage, G. Peter",
    title = "{A New Algorithm for Adaptive Multidimensional Integration}",
    reportNumber = "SLAC-PUB-1839-REV, SLAC-PUB-1839",
    doi = "10.1016/0021-9991(78)90004-9",
    journal = "J. Comput. Phys.",
    volume = "27",
    pages = "192",
    year = "1978"
}

@article{Chawdhry:2020for,
    author = "Chawdhry, Herschel A. and Czakon, Michal and Mitov, Alexander and Poncelet, Rene",
    title = "{Two-loop leading-color helicity amplitudes for three-photon production at the LHC}",
    eprint = "2012.13553",
    archivePrefix = "arXiv",
    primaryClass = "hep-ph",
    reportNumber = "Cavendish-HEP-20/15, OUTP-20-16P, P3H-20-084, TTK-20-50",
    doi = "10.1007/JHEP06(2021)150",
    journal = "JHEP",
    volume = "2021",
    number = "06",
    pages = "150",
    year = "2021"
}

@article{Binoth:2010xt,
    author = "Binoth, T. and others",
    editor = "Belanger, Genevieve and Boudjema, Fawzi and Guillet, Jean-Philippe",
    title = "{A Proposal for a Standard Interface between Monte Carlo Tools and One-Loop Programs}",
    eprint = "1001.1307",
    archivePrefix = "arXiv",
    primaryClass = "hep-ph",
    reportNumber = "SLAC-PUB-14723, FERMILAB-PUB-10-029-T, LAPTH-003-10",
    doi = "10.1016/j.cpc.2010.05.016",
    journal = "Comput. Phys. Commun.",
    volume = "181",
    pages = "1612--1622",
    year = "2010"
}

@article{Lai:2010vv,
    author = "Lai, Hung-Liang and Guzzi, Marco and Huston, Joey and Li, Zhao and Nadolsky, Pavel M. and Pumplin, Jon and Yuan, C. -P.",
    title = "{New parton distributions for collider physics}",
    eprint = "1007.2241",
    archivePrefix = "arXiv",
    primaryClass = "hep-ph",
    reportNumber = "MSUHEP-100707, SMU-HEP-10-10",
    doi = "10.1103/PhysRevD.82.074024",
    journal = "Phys. Rev. D",
    volume = "82",
    pages = "074024",
    year = "2010"
}

@article{Buckley:2014ana,
    author = {Buckley, Andy and Ferrando, James and Lloyd, Stephen and Nordstr{\"o}m, Karl and Page, Ben and R{\"u}fenacht, Martin and Sch{\"o}nherr, Marek and Watt, Graeme},
    title = "{LHAPDF6: parton density access in the LHC precision era}",
    eprint = "1412.7420",
    archivePrefix = "arXiv",
    primaryClass = "hep-ph",
    reportNumber = "GLAS-PPE-2014-05, MCNET-14-29, IPPP-14-111, DCPT-14-222",
    doi = "10.1140/epjc/s10052-015-3318-8",
    journal = "Eur. Phys. J. C",
    volume = "75",
    pages = "132",
    year = "2015"
}

@article{Agarwal:2024jyq,
    author = "Agarwal, Bakul and Heinrich, Gudrun and Jones, Stephen P. and Kerner, Matthias and Klein, Sven Yannick and Lang, Jannis and Magerya, Vitaly and Olsson, Anton",
    title = "{Two-loop amplitudes for $ t\overline{t}H $ production: the quark-initiated N$_{f}$-part}",
    eprint = "2402.03301",
    archivePrefix = "arXiv",
    primaryClass = "hep-ph",
    reportNumber = "IPPP/24/03, KA-TP-02-2024, P3H-24-007, TTK-24-03",
    doi = "10.1007/JHEP05(2024)013",
    journal = "JHEP",
    volume = "2024",
    number = "05",
    pages = "013",
    year = "2024",
    note = "[Erratum: JHEP 06, 142 (2024)]"
}

@article{Sterman:1978bi,
    author = "Sterman, George F.",
    title = "{Mass Divergences in Annihilation Processes. 1. Origin and Nature of Divergences in Cut Vacuum Polarization Diagrams}",
    reportNumber = "ITP-SB-77-69",
    doi = "10.1103/PhysRevD.17.2773",
    journal = "Phys. Rev. D",
    volume = "17",
    pages = "2773",
    year = "1978"
}

@article{Libby:1978qf,
    author = "Libby, Stephen B. and Sterman, George F.",
    title = "{Jet and Lepton Pair Production in High-Energy Lepton-Hadron and Hadron-Hadron Scattering}",
    reportNumber = "ITP-SB-78-42",
    doi = "10.1103/PhysRevD.18.3252",
    journal = "Phys. Rev. D",
    volume = "18",
    pages = "3252",
    year = "1978"
}

@article{Capatti:2019edf,
    author = "Capatti, Zeno and Hirschi, Valentin and Kermanschah, Dario and Pelloni, Andrea and Ruijl, Ben",
    title = "{Numerical Loop-Tree Duality: contour deformation and subtraction}",
    eprint = "1912.09291",
    archivePrefix = "arXiv",
    primaryClass = "hep-ph",
    doi = "10.1007/JHEP04(2020)096",
    journal = "JHEP",
    volume = "2020",
    number = "04",
    pages = "096",
    year = "2020"
}

@article{Sterman:2002qn,
    author = "Sterman, George F. and Tejeda-Yeomans, Maria E.",
    title = "{Multiloop amplitudes and resummation}",
    eprint = "hep-ph/0210130",
    archivePrefix = "arXiv",
    reportNumber = "YITP-SB-02-56",
    doi = "10.1016/S0370-2693(02)03100-3",
    journal = "Phys. Lett. B",
    volume = "552",
    pages = "48--56",
    year = "2003"
}

@article{Collins:1989gx,
    author = "Collins, John C. and Soper, Davison E. and Sterman, George F.",
    title = "{Factorization of Hard Processes in QCD}",
    eprint = "hep-ph/0409313",
    archivePrefix = "arXiv",
    reportNumber = "ITP-SB-89-31",
    doi = "10.1142/9789814503266_0001",
    journal = "Adv. Ser. Direct. High Energy Phys.",
    volume = "5",
    pages = "1--91",
    year = "1989"
}

@misc{Vermaseren:2000nd,
    author = "Vermaseren, J. A. M.",
    title = "{New features of FORM}",
    eprint = "math-ph/0010025",
    archivePrefix = "arXiv",
    month = "10",
    year = "2000"
}

@article{Nagy:2006xy,
    author = "Nagy, Zoltan and Soper, Davison E.",
    title = "{Numerical integration of one-loop Feynman diagrams for N-photon amplitudes}",
    eprint = "hep-ph/0610028",
    archivePrefix = "arXiv",
    reportNumber = "CERN-PH-TH-2006-193",
    doi = "10.1103/PhysRevD.74.093006",
    journal = "Phys. Rev. D",
    volume = "74",
    pages = "093006",
    year = "2006"
}

@article{Soper:1999xk,
    author = "Soper, Davison E.",
    title = "{Techniques for QCD calculations by numerical integration}",
    eprint = "hep-ph/9910292",
    archivePrefix = "arXiv",
    doi = "10.1103/PhysRevD.62.014009",
    journal = "Phys. Rev. D",
    volume = "62",
    pages = "014009",
    year = "2000"
}

@article{Capatti:2022tit,
    author = "Capatti, Zeno and Hirschi, Valentin and Ruijl, Ben",
    title = "{Local unitarity: cutting raised propagators and localising renormalisation}",
    eprint = "2203.11038",
    archivePrefix = "arXiv",
    primaryClass = "hep-ph",
    doi = "10.1007/JHEP10(2022)120",
    journal = "JHEP",
    volume = "2022",
    number = "10",
    pages = "120",
    year = "2022"
}

@article{Nagy:2003qn,
    author = "Nagy, Zoltan and Soper, Davison E.",
    title = "{General subtraction method for numerical calculation of one loop QCD matrix elements}",
    eprint = "hep-ph/0308127",
    archivePrefix = "arXiv",
    doi = "10.1088/1126-6708/2003/09/055",
    journal = "JHEP",
    volume = "2003",
    number = "09",
    pages = "055",
    year = "2003"
}

@article{Ma:2019hjq,
    author = "Ma, Yao",
    title = "{A Forest Formula to Subtract Infrared Singularities in Amplitudes for Wide-angle Scattering}",
    eprint = "1910.11304",
    archivePrefix = "arXiv",
    primaryClass = "hep-ph",
    reportNumber = "YITP-SB-19-34",
    doi = "10.1007/JHEP05(2020)012",
    journal = "JHEP",
    volume = "2020",
    number = "05",
    pages = "012",
    year = "2020"
}

@article{Akhoury:1978vq,
    author = "Akhoury, R.",
    title = "{Mass Divergences of Wide Angle Scattering Amplitudes}",
    reportNumber = "ITP-SB-78-43",
    doi = "10.1103/PhysRevD.19.1250",
    journal = "Phys. Rev. D",
    volume = "19",
    pages = "1250",
    year = "1979"
}

@article{Sen:1982bt,
    author = "Sen, Ashoke",
    title = "{Asymptotic Behavior of the Wide Angle On-Shell Quark Scattering Amplitudes in Nonabelian Gauge Theories}",
    reportNumber = "FERMILAB-PUB-82-066-THY, FERMILAB-PUB-82-066-T",
    doi = "10.1103/PhysRevD.28.860",
    journal = "Phys. Rev. D",
    volume = "28",
    pages = "860",
    year = "1983"
}

@article{Collins:1989bt,
    author = "Collins, John C.",
    title = "{Sudakov form-factors}",
    eprint = "hep-ph/0312336",
    archivePrefix = "arXiv",
    doi = "10.1142/9789814503266_0006",
    journal = "Adv. Ser. Direct. High Energy Phys.",
    volume = "5",
    pages = "573--614",
    year = "1989"
}

@article{Becher:2009cu,
    author = "Becher, Thomas and Neubert, Matthias",
    title = "{Infrared singularities of scattering amplitudes in perturbative QCD}",
    eprint = "0901.0722",
    archivePrefix = "arXiv",
    primaryClass = "hep-ph",
    reportNumber = "FERMILAB-PUB-09-002-T, MZ-TH-09-01",
    doi = "10.1103/PhysRevLett.102.162001",
    journal = "Phys. Rev. Lett.",
    volume = "102",
    pages = "162001",
    year = "2009",
    note = "[Erratum: Phys.Rev.Lett. 111, 199905 (2013)]"
}

@article{Feige:2014wja,
    author = "Feige, Ilya and Schwartz, Matthew D.",
    title = "{Hard-Soft-Collinear Factorization to All Orders}",
    eprint = "1403.6472",
    archivePrefix = "arXiv",
    primaryClass = "hep-ph",
    doi = "10.1103/PhysRevD.90.105020",
    journal = "Phys. Rev. D",
    volume = "90",
    number = "10",
    pages = "105020",
    year = "2014"
}

@article{Becker:2012aqa,
    author = "Becker, Sebastian and Reuschle, Christian and Weinzierl, Stefan",
    title = "{Efficiency Improvements for the Numerical Computation of NLO Corrections}",
    eprint = "1205.2096",
    archivePrefix = "arXiv",
    primaryClass = "hep-ph",
    doi = "10.1007/JHEP07(2012)090",
    journal = "JHEP",
    volume = "2012",
    number = "07",
    pages = "090",
    year = "2012"
}

@article{Gong:2008ww,
    author = "Gong, Wei and Nagy, Zoltan and Soper, Davison E.",
    title = "{Direct numerical integration of one-loop Feynman diagrams for N-photon amplitudes}",
    eprint = "0812.3686",
    archivePrefix = "arXiv",
    primaryClass = "hep-ph",
    reportNumber = "DESY-08-196",
    doi = "10.1103/PhysRevD.79.033005",
    journal = "Phys. Rev. D",
    volume = "79",
    pages = "033005",
    year = "2009"
}

@article{Becker:2012bi,
    author = "Becker, Sebastian and Weinzierl, Stefan",
    title = "{Direct numerical integration for multi-loop integrals}",
    eprint = "1211.0509",
    archivePrefix = "arXiv",
    primaryClass = "hep-ph",
    doi = "10.1140/epjc/s10052-013-2321-1",
    journal = "Eur. Phys. J. C",
    volume = "73",
    number = "2",
    pages = "2321",
    year = "2013"
}

@article{Becker:2010ng,
    author = "Becker, Sebastian and Reuschle, Christian and Weinzierl, Stefan",
    title = "{Numerical NLO QCD calculations}",
    eprint = "1010.4187",
    archivePrefix = "arXiv",
    primaryClass = "hep-ph",
    doi = "10.1007/JHEP12(2010)013",
    journal = "JHEP",
    volume = "2010",
    number = "12",
    pages = "013",
    year = "2010"
}

@misc{Anastasiou:2024xvk,
    author = "Anastasiou, Charalampos and Karlen, Julia and Sterman, George and Venkata, Aniruddha",
    title = "{Locally finite two-loop amplitudes for electroweak production through gluon fusion}",
    eprint = "2403.13712",
    archivePrefix = "arXiv",
    primaryClass = "hep-ph",
    month = "3",
    year = "2024"
}

@article{Bodwin:1984hc,
    author = "Bodwin, Geoffrey T.",
    title = "{Factorization of the Drell-Yan Cross-Section in Perturbation Theory}",
    reportNumber = "ANL-HEP-PR-84-64-REV, ANL-HEP-PR-84-64",
    journal = "Phys. Rev. D",
    volume = "31",
    pages = "2616",
    year = "1985",
    note = "[Erratum: Phys.Rev.D 34, 3932 (1986)]"
}

@article{Collins:1985ue,
    author = "Collins, John C. and Soper, Davison E. and Sterman, George F.",
    title = "{Factorization for Short Distance Hadron - Hadron Scattering}",
    reportNumber = "OITS-287",
    doi = "10.1016/0550-3213(85)90565-6",
    journal = "Nucl. Phys. B",
    volume = "261",
    pages = "104--142",
    year = "1985"
}

@article{Aguilera-Verdugo:2019kbz,
    author = "Aguilera-Verdugo, J. Jes\'us and Driencourt-Mangin, F\'elix and Plenter, Judith and Ram\'\i{}rez-Uribe, Selomit and Rodrigo, Germ\'an and Sborlini, Germ\'an F. R. and Torres Bobadilla, William J. and Tracz, Szymon",
    title = "{Causality, unitarity thresholds, anomalous thresholds and infrared singularities from the loop-tree duality at higher orders}",
    eprint = "1904.08389",
    archivePrefix = "arXiv",
    primaryClass = "hep-ph",
    reportNumber = "IFIC/19-22",
    doi = "10.1007/JHEP12(2019)163",
    journal = "JHEP",
    volume = "2019",
    number = "12",
    pages = "163",
    year = "2019"
}

@article{JesusAguilera-Verdugo:2020fsn,
    author = "Jes\'us Aguilera-Verdugo, J. and Hern\'andez-Pinto, Roger J. and Rodrigo, Germ\'an and Sborlini, German F. R. and Torres Bobadilla, William J.",
    title = "{Mathematical properties of nested residues and their application to multi-loop scattering amplitudes}",
    eprint = "2010.12971",
    archivePrefix = "arXiv",
    primaryClass = "hep-ph",
    reportNumber = "IFIC/20-30; DESY 20-172; MPP-2020-184",
    doi = "10.1007/JHEP02(2021)112",
    journal = "JHEP",
    volume = "2021",
    number = "02",
    pages = "112",
    year = "2021"
}

@article{Ramirez-Uribe:2022sja,
    author = "Ram\'\i{}rez-Uribe, Selomit and Hern\'andez-Pinto, Roger Jos\'e and Rodrigo, Germ\'an and Sborlini, German F. R.",
    title = "{From Five-Loop Scattering Amplitudes to Open Trees with the Loop-Tree Duality}",
    eprint = "2211.03163",
    archivePrefix = "arXiv",
    primaryClass = "hep-ph",
    reportNumber = "IFIC/20-30",
    doi = "10.3390/sym14122571",
    journal = "Symmetry",
    volume = "14",
    number = "12",
    pages = "2571",
    year = "2022"
}

@article{Runkel:2019yrs,
    author = "Runkel, Robert and Sz\H{o}r, Zolt\'an and Vesga, Juan Pablo and Weinzierl, Stefan",
    title = "{Causality and loop-tree duality at higher loops}",
    eprint = "1902.02135",
    archivePrefix = "arXiv",
    primaryClass = "hep-ph",
    doi = "10.1103/PhysRevLett.122.111603",
    journal = "Phys. Rev. Lett.",
    volume = "122",
    number = "11",
    pages = "111603",
    year = "2019",
    note = "[Erratum: Phys.Rev.Lett. 123, 059902 (2019)]"
}

@article{Aguilera-Verdugo:2020set,
    author = "Aguilera-Verdugo, J. Jesus and Driencourt-Mangin, Felix and Hern\'andez-Pinto, Roger J. and Plenter, Judith and Ramirez-Uribe, Selomit and Renteria Olivo, Andres E. and Rodrigo, German and Sborlini, German F. R. and Torres Bobadilla, William J. and Tracz, Szymon",
    title = "{Open Loop Amplitudes and Causality to All Orders and Powers from the Loop-Tree Duality}",
    eprint = "2001.03564",
    archivePrefix = "arXiv",
    primaryClass = "hep-ph",
    reportNumber = "IFIC/20-02",
    doi = "10.1103/PhysRevLett.124.211602",
    journal = "Phys. Rev. Lett.",
    volume = "124",
    number = "21",
    pages = "211602",
    year = "2020"
}

@article{Aguilera-Verdugo:2020kzc,
    author = "Aguilera-Verdugo, J. Jesus and Hernandez-Pinto, Roger J. and Rodrigo, German and Sborlini, German F. R. and Torres Bobadilla, William J.",
    title = "{Causal representation of multi-loop Feynman integrands within the loop-tree duality}",
    eprint = "2006.11217",
    archivePrefix = "arXiv",
    primaryClass = "hep-ph",
    reportNumber = "IFIC/20-27",
    doi = "10.1007/JHEP01(2021)069",
    journal = "JHEP",
    volume = "2021",
    number = "01",
    pages = "069",
    year = "2021"
}

@article{Ramirez-Uribe:2020hes,
    author = "Ram\'\i{}rez-Uribe, Selomit and Hern\'andez-Pinto, Roger J. and Rodrigo, German and Sborlini, Germ\'an F. R. and Torres Bobadilla, William J.",
    title = "{Universal opening of four-loop scattering amplitudes to trees}",
    eprint = "2006.13818",
    archivePrefix = "arXiv",
    primaryClass = "hep-ph",
    reportNumber = "IFIC/20-29",
    doi = "10.1007/JHEP04(2021)129",
    journal = "JHEP",
    volume = "2021",
    number = "04",
    pages = "129",
    year = "2021"
}

@article{Sborlini:2021owe,
    author = "Sborlini, German F. R.",
    title = "{Geometrical approach to causality in multiloop amplitudes}",
    eprint = "2102.05062",
    archivePrefix = "arXiv",
    primaryClass = "hep-ph",
    reportNumber = "DESY-21-017, DESY 21-017",
    doi = "10.1103/PhysRevD.104.036014",
    journal = "Phys. Rev. D",
    volume = "104",
    number = "3",
    pages = "036014",
    year = "2021"
}

@article{TorresBobadilla:2021ivx,
    author = "Torres Bobadilla, William J.",
    title = "{Loop-tree duality from vertices and edges}",
    eprint = "2102.05048",
    archivePrefix = "arXiv",
    primaryClass = "hep-ph",
    reportNumber = "MPP-2021-14",
    doi = "10.1007/JHEP04(2021)183",
    journal = "JHEP",
    volume = "2021",
    number = "04",
    pages = "183",
    year = "2021"
}

@article{Benincasa:2021qcb,
    author = "Benincasa, Paolo and Bobadilla, William J. Torres",
    title = "{Physical representations for scattering amplitudes and the wavefunction of the universe}",
    eprint = "2112.09028",
    archivePrefix = "arXiv",
    primaryClass = "hep-th",
    reportNumber = "MPP-2021-186",
    doi = "10.21468/SciPostPhys.12.6.192",
    journal = "SciPost Phys.",
    volume = "12",
    number = "6",
    pages = "192",
    year = "2022"
}

@article{Kromin:2022txz,
    author = "Kromin, Sascha and Schwanemann, Niklas and Weinzierl, Stefan",
    title = "{Amplitudes within causal loop-tree duality}",
    eprint = "2208.01060",
    archivePrefix = "arXiv",
    primaryClass = "hep-th",
    reportNumber = "MITP/22-071",
    doi = "10.1103/PhysRevD.106.076006",
    journal = "Phys. Rev. D",
    volume = "106",
    number = "7",
    pages = "076006",
    year = "2022"
}

@misc{Capatti:2020ytd,
    author = "Capatti, Zeno and Hirschi, Valentin and Kermanschah, Dario and Pelloni, Andrea and Ruijl, Ben",
    title = "{Manifestly Causal Loop-Tree Duality}",
    eprint = "2009.05509",
    archivePrefix = "arXiv",
    primaryClass = "hep-ph",
    month = "9",
    year = "2020"
}

@article{Buchta:2015wna,
    author = "Buchta, Sebastian and Chachamis, Grigorios and Draggiotis, Petros and Rodrigo, German",
    title = "{Numerical implementation of the loop\textendash{}tree duality method}",
    eprint = "1510.00187",
    archivePrefix = "arXiv",
    primaryClass = "hep-ph",
    reportNumber = "IFIC-15-69",
    doi = "10.1140/epjc/s10052-017-4833-6",
    journal = "Eur. Phys. J. C",
    volume = "77",
    number = "5",
    pages = "274",
    year = "2017"
}

@misc{Kilian:2009wy,
    author = "Kilian, Wolfgang and Kleinschmidt, Tobias",
    title = "{Numerical Evaluation of Feynman Loop Integrals by Reduction to Tree Graphs}",
    eprint = "0912.3495",
    archivePrefix = "arXiv",
    primaryClass = "hep-ph",
    reportNumber = "IPPP-09-97, DCTP-09-194, EDINBURGH-2009-17, SI-HEP-2009-16",
    month = "12",
    year = "2009"
}

@misc{Anastasiou:2025cvy,
    author = "Anastasiou, Charalampos and Karlen, Julia and Sahoo, Roshni and Sterman, George and Vicini, Matilde",
    title = "{General finite two-loop amplitude integrand for photoproduction in quark annihilation}",
    eprint = "2509.07805",
    archivePrefix = "arXiv",
    primaryClass = "hep-ph",
    month = "9",
    year = "2025"
}

@article{Kuipers:2013pba,
    author = "Kuipers, J. and Ueda, T. and Vermaseren, J. A. M.",
    title = "{Code Optimization in FORM}",
    eprint = "1310.7007",
    archivePrefix = "arXiv",
    primaryClass = "cs.SC",
    reportNumber = "NIKHEF-2013-036, TTP13-031, SFB-CPP-13-80",
    doi = "10.1016/j.cpc.2014.08.008",
    journal = "Comput. Phys. Commun.",
    volume = "189",
    pages = "1--19",
    year = "2015"
}

@misc{symbolica,
  author = {Ruijl, Ben},
  title = {{Symbolica}},
  year = {2025},
  note = {Version 0.18.0},
  doi = {10.5281/zenodo.17054381},
  url = {https://doi.org/10.5281/zenodo.17054381}
}

@article{Kleiss:1985gy,
    author = "Kleiss, R. and Stirling, W. James and Ellis, S. D.",
    title = "{A New Monte Carlo Treatment of Multiparticle Phase Space at High-energies}",
    reportNumber = "CERN-TH-4299/85",
    doi = "10.1016/0010-4655(86)90119-0",
    journal = "Comput. Phys. Commun.",
    volume = "40",
    pages = "359",
    year = "1986"
}

@article{Haindl:2025jte,
    author = "Haindl, Rayan",
    title = "{Local form factor subtraction for three-loop QCD corrections to electroweak production in quark-antiquark annihilation}",
    eprint = "2502.00572",
    archivePrefix = "arXiv",
    primaryClass = "hep-ph",
    doi = "10.1007/JHEP06(2025)257",
    journal = "JHEP",
    volume = "2025",
    number = "06",
    pages = "257",
    year = "2025"
}

@article{Ohl:1998jn,
    author = "Ohl, Thorsten",
    title = "{Vegas revisited: Adaptive Monte Carlo integration beyond factorization}",
    eprint = "hep-ph/9806432",
    archivePrefix = "arXiv",
    reportNumber = "IKDA-98-15",
    doi = "10.1016/S0010-4655(99)00209-X",
    journal = "Comput. Phys. Commun.",
    volume = "120",
    pages = "13--19",
    year = "1999"
}

@article{Kleiss:1994qy,
    author = "Kleiss, Ronald and Pittau, Roberto",
    title = "{Weight optimization in multichannel Monte Carlo}",
    eprint = "hep-ph/9405257",
    archivePrefix = "arXiv",
    reportNumber = "NIKHEF-H-94-17, INLO-PUB-4-94",
    doi = "10.1016/0010-4655(94)90043-4",
    journal = "Comput. Phys. Commun.",
    volume = "83",
    pages = "141--146",
    year = "1994"
}

@article{Caswell:1981ek,
    author = "Caswell, William E. and Kennedy, A. D.",
    title = "{A SIMPLE APPROACH TO RENORMALIZATION THEORY}",
    reportNumber = "MdDP-PP-81-107",
    doi = "10.1103/PhysRevD.25.392",
    journal = "Phys. Rev. D",
    volume = "25",
    pages = "392",
    year = "1982"
}

@article{Gomes:1974cr,
    author = "Gomes, M. and Lowenstein, J. H. and Zimmermann, W.",
    title = "{Generalization of the Momentum-Space Subtraction Procedure for Renormalized Perturbation Theory}",
    reportNumber = "NYU/TR11/74",
    doi = "10.1007/BF01608388",
    journal = "Commun. Math. Phys.",
    volume = "39",
    pages = "81",
    year = "1974"
}

@article{Herzog:2017jgk,
    author = "Herzog, Franz",
    title = "{Zimmermann's forest formula, infrared divergences and the QCD beta function}",
    eprint = "1711.06121",
    archivePrefix = "arXiv",
    primaryClass = "hep-ph",
    doi = "10.1016/j.nuclphysb.2017.11.011",
    journal = "Nucl. Phys. B",
    volume = "926",
    pages = "370--380",
    year = "2018"
}

@misc{havana,
  author = {Ruijl, Ben and Kermanschah, Dario},
  title = {{Havana}},
  year = {2025},
  howpublished = {Original implementation at \url{https://github.com/benruijl/havana}; calculations use the \href{https://github.com/kermanschah/havana/tree/91292440a1e55074ed546ead8c35e228aadd5447}{Kermanschah fork at revision \texttt{9129244}}},
  note = {Version 0.1.6}
}

@article{Badger:2024sqv,
    author = "Badger, Simon and Hartanto, Heribertus Bayu and Wu, Zihao and Zhang, Yang and Zoia, Simone",
    title = "{Two-loop amplitudes for $\mathcal{O}(\alpha_s^2)$ corrections to $W\gamma\gamma$ production at the LHC}",
    eprint = "2409.08146",
    archivePrefix = "arXiv",
    primaryClass = "hep-ph",
    doi = "10.1007/JHEP12(2024)221",
    journal = "JHEP",
    volume = "2024",
    number = "12",
    pages = "221",
    year = "2024"
}

@article{Badger:2024dxo,
    author = "Badger, Simon and Becchetti, Matteo and Brancaccio, Colomba and Hartanto, Heribertus Bayu and Zoia, Simone",
    title = "{Numerical evaluation of two-loop QCD helicity amplitudes for $ gg\to t\overline{t}g $ at leading colour}",
    eprint = "2412.13876",
    archivePrefix = "arXiv",
    primaryClass = "hep-ph",
    reportNumber = "ZU-TH-65/24",
    doi = "10.1007/JHEP03(2025)070",
    journal = "JHEP",
    volume = "2025",
    number = "03",
    pages = "070",
    year = "2025"
}

@article{Badger:2025ljy,
    author = "Badger, Simon and Becchetti, Matteo and Brancaccio, Colomba and Czakon, Micha{\l} and Hartanto, Heribertus Bayu and Poncelet, Rene and Zoia, Simone",
    title = "{Double virtual QCD corrections to $ t\overline{t} $+jet production at the LHC}",
    eprint = "2511.11424",
    archivePrefix = "arXiv",
    primaryClass = "hep-ph",
    reportNumber = "P3H-25-092, TTK-25-38, ZU-TH 74/25, IFJPAN-IV-2025-23",
    doi = "10.1007/JHEP05(2026)044",
    journal = "JHEP",
    volume = "2026",
    number = "05",
    pages = "044",
    year = "2026"
}

@misc{Becchetti:2026yxl,
    author = "Becchetti, Matteo and Canko, Dhimiter and Chen, Xiang and Chestnov, Vsevolod and Delto, Maximilian and Ditsch, Sara and Peraro, Tiziano and Pozzoli, Mattia and Savoini, Chiara and Zoia, Simone",
    title = "{Two-loop QCD amplitudes for $t\bar{t}W$ production at the LHC in the leading-colour approximation}",
    eprint = "2608.03746",
    archivePrefix = "arXiv",
    primaryClass = "hep-ph",
    reportNumber = "ZU-TH 27/26, TUM-HEP-1610/26, MPP-2026-132, CERN-TH-2026-183",
    month = "8",
    year = "2026"
}

@article{Driencourt-Mangin:2019aix,
    author = "Driencourt-Mangin, Felix and Rodrigo, Germ{\'a}n and Sborlini, Germ{\'a}n F. R. and Torres Bobadilla, William J.",
    title = "{Universal four-dimensional representation of $H \to \gamma \gamma$ at two loops through the Loop-Tree Duality}",
    eprint = "1901.09853",
    archivePrefix = "arXiv",
    primaryClass = "hep-ph",
    reportNumber = "IFIC/18-31, TIF-UNIMI-2018-6",
    doi = "10.1007/JHEP02(2019)143",
    journal = "JHEP",
    volume = "2019",
    number = "02",
    pages = "143",
    year = "2019"
}

@article{Rios-Sanchez:2024xtv,
    author = "Rios-Sanchez, Jose and Sborlini, German",
    title = "{Toward multiloop local renormalization within causal loop-tree duality}",
    eprint = "2402.13995",
    archivePrefix = "arXiv",
    primaryClass = "hep-th",
    doi = "10.1103/PhysRevD.109.125004",
    journal = "Phys. Rev. D",
    volume = "109",
    number = "12",
    pages = "125004",
    year = "2024"
}

@article{Hartanto:2019uvl,
    author = "Hartanto, Heribertus Bayu and Badger, Simon and Br{\o}nnum-Hansen, Christian and Peraro, Tiziano",
    title = "{A numerical evaluation of planar two-loop helicity amplitudes for a W-boson plus four partons}",
    eprint = "1906.11862",
    archivePrefix = "arXiv",
    primaryClass = "hep-ph",
    reportNumber = "IPPP/19/56, ZU-TH 33/19",
    doi = "10.1007/JHEP09(2019)119",
    journal = "JHEP",
    volume = "2019",
    number = "09",
    pages = "119",
    year = "2019"
}

@article{Badger:2021nhg,
    author = "Badger, Simon and Hartanto, Heribertus Bayu and Zoia, Simone",
    title = "{Two-Loop QCD Corrections to $Wb\bar b$ Production at Hadron Colliders}",
    eprint = "2102.02516",
    archivePrefix = "arXiv",
    primaryClass = "hep-ph",
    reportNumber = "CAVENDISH-HEP-21/01",
    doi = "10.1103/PhysRevLett.127.012001",
    journal = "Phys. Rev. Lett.",
    volume = "127",
    number = "1",
    pages = "012001",
    year = "2021"
}

@article{Abreu:2021asb,
    author = "Abreu, S. and Febres Cordero, F. and Ita, H. and Klinkert, M. and Page, B. and Sotnikov, V.",
    title = "{Leading-color two-loop amplitudes for four partons and a W boson in QCD}",
    eprint = "2110.07541",
    archivePrefix = "arXiv",
    primaryClass = "hep-ph",
    reportNumber = "CERN-TH-2021-156, FR-PHENO-2021-12, MPP-2021-181",
    doi = "10.1007/JHEP04(2022)042",
    journal = "JHEP",
    volume = "2022",
    number = "04",
    pages = "042",
    year = "2022"
}

@article{Badger:2024mir,
    author = "Badger, Simon and Hartanto, Heribertus Bayu and Poncelet, Rene and Wu, Zihao and Zhang, Yang and Zoia, Simone",
    title = "{Full-colour double-virtual amplitudes for associated production of a Higgs boson with a bottom-quark pair at the LHC}",
    eprint = "2412.06519",
    archivePrefix = "arXiv",
    primaryClass = "hep-ph",
    reportNumber = "IFJPAN-IV-2024-14, USTC-ICTS/PCFT-24-55, ZU-TH-63/24",
    doi = "10.1007/JHEP03(2025)066",
    journal = "JHEP",
    volume = "2025",
    number = "03",
    pages = "066",
    year = "2025"
}

@article{DeLaurentis:2025dxw,
    author = "De Laurentis, Giuseppe and Ita, Harald and Page, Ben and Sotnikov, Vasily",
    title = "{Compact two-loop QCD corrections for Vjj production in proton collisions}",
    eprint = "2503.10595",
    archivePrefix = "arXiv",
    primaryClass = "hep-ph",
    reportNumber = "PSI-PR-24-29, ZU-TH 14/25",
    doi = "10.1007/JHEP06(2025)093",
    journal = "JHEP",
    volume = "2025",
    number = "06",
    pages = "093",
    year = "2025"
}

@article{DeLaurentis:2026brm,
    author = "De Laurentis, Giuseppe and Ita, Harald and Kuschke, Viktor and Ruf, Michael and Sotnikov, Vasily",
    title = "{Two-loop leading-color QCD corrections for Higgs plus two-jet production in the heavy-top limit}",
    eprint = "2605.04009",
    archivePrefix = "arXiv",
    primaryClass = "hep-ph",
    reportNumber = "ZU-TH 18/26, SLAC-PUB-260429",
    doi = "10.1007/JHEP07(2026)251",
    journal = "JHEP",
    volume = "2026",
    number = "07",
    pages = "251",
    year = "2026"
}

\end{document}